\documentclass{elsart}

\usepackage{amssymb}

\usepackage{graphicx}
\usepackage{epsfig}
\usepackage{amsfonts}
\begin{document}

\begin{frontmatter}



\title{Coupled Effects in Quantum Dot Nanostructures with Nonlinear
Strain and Bridging Modelling Scales}


\author{Roderick Melnik and Roy Mahapatra}

\address{Mathematical Modelling \& Computational Sciences, \\
Wilfrid Laurier University, Waterloo Campus, \\
75 University Avenue West, Waterloo, ON, Canada N2L 3C5}

\begin{abstract}
% Text of abstract
We demonstrate that the conventional application of linear models to
the analysis of optoelectromechanical properties of nanostructures
in bandstructure engineering could be inadequate. Such linear models
are usually derived from the traditional bottom-up approach applied
to the analysis of
 nanostructure properties. At the same time, in the hierarchy of mathematical
models for semiconductor device modelling constructed on the basis
of the top-down approach, we deal predominantly with models where
nonlinearity is essential. In this contribution, we analyze these
two fundamental approaches in bridging the scales in mathematical
models for the description of optoelectromechanical properties of
nanostructures. The focus of the present paper is on a model based
on the coupled Schrodinger-Poisson system where we account
consistently for the piezoelectric effect and analyze the influence
of different nonlinear terms in strain components. The examples
given in this paper show that the piezoelectric effect contributions
are essential and have to be accounted for with fully coupled
models. While in structural applications of piezoelectric materials
at larger scales, the minimization of the full electromechanical
energy is now a routine in many engineering applications, in
bandstructure engineering conventional approaches are still based on
linear models with minimization of uncoupled, purely elastic energy
functionals with respect to displacements. Generalizations of the
existing models for bandstructure calculations are presented in this
paper in the context of coupled effects.
\end{abstract}

\end{frontmatter}

% main text
\section{Introduction}
\label{}

In low-dimensional semiconductor nanostructures (LDSN) the motion of electrons can be
confined spatially, from one, two, and even three spatial directions. In the latter case,
such nanostructures are known as quantum dots and often termed by physicists as 0-dimensional
structures, reflecting the fact that the motion of carriers is constrained from all three
spatial directions. These structures have been receiving an increasing interest due to new
technological advances and fascinating applications they offer. Indeed, they can be used as
biological tags in cell biology and biomedicine, be used in constructing quantum bits for
quantum computing, and be applied in a wide range of more traditional structure- and
device-like applications, including photodectors, laser-based emitters, etc.

While in many such applications the focus is on optical properties of future devices, it is
important to remember that the formation of LDSNs, and in particular quantum dots, is a
competition between the surface energy in the structure and strain energy. Hence, mechanical
properties are essential in designing quantum-dot-based devices and structures. Further, many
quantum dot structures have a well pronounced piezoelectric effect which does contribute to
their overall properties in a non-trivial manner. These coupled electromechanical effects
will become increasingly important for the current and future applications of such
nanostructures. In designing stable strained nanostructures, computational modelling provides
a major tool for predicting their optoelectromechanical properties.

During the last decade, the attention of the science and engineering community to the
influence of strain effects on quantum mechanical properties of LDSNs has been growing
rapidly \cite{Grundmann1995,Andreev2000,Johnson2003,Fonoberov2003,Kuo2006}. A majority of the
published works were focusing on strain effects only, without taking into account
electromechanical interactions due to the piezoelectric effect. Those authors who did account
for the piezoeffect based their considerations on the minimization of uncoupled, purely
elastic energy functionals with respect to displacements. Under this approach, the Maxwell
equation for piezoelectric solids and the equations of elasticity were effectively solved in
 either uncoupled or semicoupled manner. Pan with his collaborators \cite{Pan2002,Jogai2003} were the
first who have attracted the attention of the bandstructure engineering community to the
importance of coupled effects. Based on his semi-analytical Green's function approach applied
to an idealized half-space structure \cite{Pan2002}, he demonstrated that only the fully
coupled model can lend a reliable prediction. Further, based on a combination of analytical
(for the 1D case) and numerical (for the 2D case) techniques, the idea was generalized to the
device level \cite{Jogai2003}, but no details of the numerical procedure were given. All the
 above studies were based on the linear theory only. Furthermore, the nature of the analyzed problem
allowed a number of simplifications, including those related to the
wetting layer. The inclusion of the wetting layer in a consistent
manner not only increase the computational complexity of the problem
in several times due to different spatial scales, but may also
require the formulation of non-trivial boundary conditions
\cite{Melnik2004}.

In this paper, we base our consideration on the coupled
Schrodinger-Poisson model where we account consistently for the
piezoelectric effect and analyze the influence of different
nonlinear terms in strain components. We structure the paper as
follows. In Section 2, we analyze two fundamental approaches in
bridging the scales in mathematical models for the description of
optoelectromechanical properties of nanostructures. In Section 3, we
provide the core model for the description of coupled
electromechanical interactions in piezoelectric semiconductor
solids. The model is exemplified for hexagonal (WZ) and cubic (ZB)
materials used in our computational experiments. In Section 4, we
give details of the model for bandstructure calculations, focusing
on the ${\bf k} \cdot {\bf p}$ approximation as a convenient
framework for incorporating strain and piezoelectric effects. A
general procedure for modelling quantum dot nanostructures, based on
the variational formulation of the problem, is outlined in Section
5. In Section 6 we provide details of numerical experiments
demonstrating the influence of the piezoelectric effect and
analyzing contributions of nonlinear terms in strain components.
Conclusions are given in Section 7.

%%%%%%%%%%%%%%%%%%%%%%%%%%%%%%%%%%%%%%%

\section{Waves propagation in anisotropic media:
applying experience from solid and fluid mechanics to bandstructure
engineering}

Waves propagation in anisotropic media has always been a topic in
the heart of scientific inquiries and a source of new ideas for
engineers. This topic is of immense practical importance in the
context of both solid and fluid mechanics. Purely elastic waves in
shells and other structures have been studied in the context of
fluid-solid interactions at least since the late 1950ies, providing
 many meaningful examples where coupling effects become essential.
Around the same time, an increasing interest to
 coupled problems was also generated by the
analysis of dynamic thermal stresses in structures, in particular
 by the famous Danilovskaya problem, first formulated in 1950. Since then, coupled
effects, in particular in anisotropic materials, have continued to
fuel interest to their studies due to both, theoretical challenges
and an increasing range of practical applications.

In what follows, we will focus on the coupling between electric and
mechanical fields in low-dimensional semiconductor nanostructures.
The core of the mathematical models dealing with this sort of
coupling contains the equations for coupled electromechanical motion
of piezoelectric solids. The experience accumulated in mechanics of
solids in solving such equations becomes now invaluable in the area
 known as bandstructure engineering and in the modelling semiconductor
quantum structures in general. At the same time, the experience
accumulated in fluid dynamics applications is equally important in
this area. Indeed, the hydrodynamic approach in the analysis of
semiconductor devices, superlattices, and other semiconductor
structures has been an important tool in semiconductor modelling for
a number of years. Fluid-dynamics-like hydrodynamic approximations
have been widely utilized in this area and more recently several
their extensions have been proposed to account for quantum effects.
 The importance of
 such semiconductor systems as quantum wells, wires, dots, and
 superlattices \cite{Willatzen2004MCS} will continue to grow in
 nanoscale electronics, photonics, and bioengineering.  New
 technological advances in applications of these structures require
 to have a fresh look at
 their modelling aspects, in particular in the context of their optoelectromechanical
 properties. While strain effects are fundamental to such properties, in the bandstructure
 engineering literature their influence is still typically analyzed with simplified linear
 models based on the minimization of uncoupled, purely elastic energy functionals with
 respect to displacements. The applicability of such models is limited as coupled effects
 related, e.g., to built-in spontaneous and piezoelectric polarization become essential. New
 models accounting for these effects need to be developed.

The modelling experience accumulated in both mechanics of solids and
fluid mechanics can help in achieving this task. To get started,
 note that in semiconductor systems we are dealing with, both
classical and quantum effects are interlinked, and
 the analysis of such systems and the choice of modelling tools depend
  critically on the spatio-temporal
 scales required for specific applications. We can start constructing a model for
 the analysis of such systems from the
 fundamental quantum level
 by specifying the Hamiltonian of the system within the Schrodinger framework. However,
 then we should incorporate additional effects, pronounced at larger scales,
  such as piezoelectric, into the
obtained
 approximate model. This is {\em the bottom-up approach} to modelling
  semiconductors, applied actively today for the analysis of nanostructures.
  Alternatively, we can
 attempt to carry out some physics-based averaging right from the beginning, applying
  {\em the top-down approach} to modelling semiconductors.
 A good example,
  clarifying conceptually the applicability and limitations of these approaches,
    can be provided by considering models
   for superlattices. Based on the underlying physical assumptions, there are two major classes
   of such structures, classical and quantum. These structures have additional periodicity on
   a scale larger than atomic. The idea of creating quantum superlattices
 is due to L. Keldysh (1962).  Experimentalists reported the creation of such objects
  about a decade later (L. Esaki, 1970; Zh. Alferov et al, 1971). It has been the domain of
  solid state physics where tools of solid mechanics are essential and well established.
  On the other hand, the idea of creating classical semiconductor superlattices
was originated from a fluid mechanics analogy. It is well known
(e.g., \cite{Bass1997} and references therein) that a fluid under
gravity can reach its equilibrium if its temperature depends on the
height only. If the temperature gradient, directed down, exceeds
certain critical value, we observe a free convection of the fluid.
If we assume that this process takes place between two infinitely
long horizontal planes heated to different temperatures (temperature
of the lower plane is higher), under a sufficiently high temperature
gradient the fluid becomes unstable and we observe stationary
convective motion. Due to the underlying assumptions, in the
horizontal plane the motion is expected to be periodic. Based on
this hydrodynamic analogy, a similar idea was proposed in early
1970ies in the context of carrier motion in semiconductors, where
the role of gravitational field could be played by the electric
field with heating produced by, e.g., light (\cite{Bass1997} and
references
 therein). The analogy between wave phenomena in classical superlattices
 and the behaviour of wave functions of an electron moving in a periodic potential field of a
 quantum superlattice can be exploited when developing a hierarchy of models for the analysis
 of semiconductor structures. In both cases, we have an additional periodicity of the structure.
 However, the difference between these two cases lies in the fact
 that while for classical superlattices such an additional periodicity leads to the
 quantization of wave energy, in quantum superlattices it leads to the quantization of
 carrier energy. If the period of the potential field exceeds the length of free carrier
 runs, so that on this specific spatio-temporal scale carriers will not be affected by the
 action of the additional periodic field, semiconductor superlattices
  behave like classical structures. In the latter case,
  many (fluid-mechanics analogy based) techniques developed
 for semiconductor
 device modelling at
 sub-micron scales can often be applied. The study of quantum superlattices
 requires more fundamental approaches to account for atomic scales.

Note that already at the classical level we have to construct a
multiscale hierarchy of the models. Indeed, the standard
drift-diffusion approximation may not be an appropriate modelling
tool even for classical superlattices, while hydrodynamic and
kinetic models provide quite useful tools
 for the analysis of such structures. Based on relaxation time
 approximations, a classification of the
 hierarchy of mathematical models for these structures was discussed
  in \cite{Melnik2000MS,Melnik2000H}. At the top of this hierarchy is
   the Liouville equation
 framework which leads to substantial difficulties in
 practical realization of this approach. Hence, most practical
 approaches stem from the kinetic-type models
 such as the semi-classical Boltzmann equation. In these models, scattering of carriers on each
 other is not essential. However, the scattering of carriers on imperfections of the lattice
 plays the dominant role. Hence, charge carriers under this approach cannot be considered as
 an independent thermodynamical system. Based on the moment methodology or the Hilbert
 expansion method, a range of macroscopic models can be derived, among which
 hydrodynamic-type
 models play a prominent role. In such models the electron-hole "plasma" can be considered
 as an almost independent
 thermodynamical system that only weakly interacts with the crystal lattice. This group of
 models as well as quasi-hydrodynamic models can account for
  non-equilibrium and non-local behaviour of semiconductor carrier "plasma".
  Further details of the developed computational techniques for such models can be found in
 \cite{Melnik2000H,Melnik2000MS,Melnik2000MCS}. We note
 that the equations we deal with in such situations are similar to, but differ from,
  the hydrodynamic equations of fluid mechanics. They consist
 of the Poisson equation, equations of continuity (for carrier concentrations) and energy transfer.
 Furthermore, re-distributions of charge carriers lead to an additional field, a phenomenon
 absent in the fluid mechanics. Attempts to apply these types of models
  to other semiconductor
 structures and devices at smaller scales
  have led recently to the development of extended hydrodynamic
  models that should incorporate quantum
  corrections \cite{Rudan2005}. The important observation is that all
   the models we have discussed above within the top-down approach are intrinsically nonlinear.

At the other end of the spectrum of model hierarchy are the models
developed with the bottom-up approach. Surprisingly, up to date  the
majority of research efforts in this area has been concentrated on
linear models. In what follows, we focus on the analysis of quantum
dot
 structures and show that the conventional approaches to the analysis of these structures
 based on linear models need to be augmented to account for coupled nonlinear effects.

%%%%%%%%%%%%%%%%%%%%%%%%%%%%%%%%%%%%%%%%

\section{Coupled electromechanical interactions in quantum dot nanostructures}

Mechanical effects profoundly influence electronic and optical
properties of the nanostructures. Two points should be mentioned in
this context. Firstly, we note that the key to intrinsic properties
of quantum dot structures lies with strain effects arising from
lattice mismatch. Following \cite{Ipatova1993}, where the authors
started their reasoning from the total Helmholtz free energy
function, we assume that there is the local equilibrium value of the
lattice constant. Hence, the
 lattice mismatch can be incorporated in the models for bandstructure
calculations by defining the strain associated with it as a mismatch
between two material layers
\begin{eqnarray}
\varepsilon_{\rm m} = (a^0 - a({\bf r}))/a^0, \label{eq1}
\end{eqnarray}
and by accounting for it in the strain-displacement relationships. In (\ref{eq1}), $a({\bf
r})$ and $a^0$ are lattice constants of two different material layers, respectively, while
${\bf r}$ is the position vector responsible for tracking the interface. This aspect of
mechanical effect contributions has been actively incorporated into the theory of
bandstructure calculations  since early 1970s, starting from fundamental works by Pikus, Bir,
Rasba, Sheka and many others \cite{Bir1974}. Secondly, semiconductors are piezoelectric
materials and the piezoelectric effect contributions to the overall properties cannot be
ignored in bandstructure engineering, neither for hexagonal (wurtzite) structures (often due
to the principle strain components) nor for cubic (zinc-blende) structures (often due to the
shear strain components). Piezoelectrics represent anisotropic media and wave interactions in
such media have been a topic of immense practical importance. While purely elastic waves in
structures have been studied intensively
 for many decades, the study of coupled electromechanical interactions in anisotropic materials is of more
recent origin. One reason for that lies with the fact that dealing with problems of coupled
electroelasticity usually requires the development and implementation of effective numerical
techniques \cite{Melnik2004ZPiezo}. Hence, the experience that has been accumulated in
solving problems of coupled electroelasticity for piezoelectric structures becomes invaluable
in the area of bandstructure engineering and modelling semiconductor quantum structures. As
in the other areas where modelling piezoelectric solids is an essential component, we
consider the following general model (e.g., \cite{Melnik2004ZPiezo}), describing coupled
electromechanical interactions in the Cartesian system of coordinates ${\bf x} =(x_1, x_2,
x_3)^T$:
\begin{eqnarray}
\rho \frac{\partial^2 {\bf u}}{\partial t^2} = \nabla \cdot \mbox{\boldmath$\sigma$} + {\bf
F}, \quad {\rm div} {\bf D} = {\bf G}, \quad {\bf E}= - \nabla \varphi, \label{eq2}
\end{eqnarray}
where ${\bf u} = (u_1, u_2, u_3)^T$ is the displacement vector, $\mbox{\boldmath$\sigma$} =
(\sigma_{ij})$ is the stress, $\rho$ is the density of the piezoelectric material, ${\bf F}$
and ${\bf G}$ are body and electric forces on the piezoelectric, if any, ${\bf E}$ and ${\bf
D}$ are the electric field and electric displacement, and $\varphi$ is the electrostatic
potential. The conventional procedure applied in modelling LDSNs is based on the minimization
of the purely elastic functionals (e.g., \cite{Fonoberov2003}), rather than on the solution
of the fully coupled problem. The effect of coupling has been analyzed rigorously in a
general setting in \cite{Melnik1998,Melnik2000,Melnik2004ZPiezo} (see also references
therein), while in the context of nanostructure modelling it has recently been demonstrated
that such an effect could be quite substantial \cite{Pan2002,Jogai2003}. The core component
of our model for analyzing the properties of quantum dot nanostructures will be the
equilibrium equations of the coupled theory of electroelasticity which are simplified in this
case to
\begin{eqnarray}
{\partial \sigma_{ij}}/{\partial x_j} =0, \quad {\rm div} {\bf D} =0, \label{eq3}
\end{eqnarray}
where the coordinate subindeces in the Timoshenko-Karman notions are obtained by changing $1
\rightarrow x$, $2 \rightarrow y$, $z \rightarrow 3$ in the tensorial representation above.
The type of coupling between the mechanical and electric fields is determined by the type of
the crystallographic symmetry of the material. In particular, for the WZ semiconductors we
have
\begin{eqnarray}
&&
 \sigma_{xx} = c_{11} \varepsilon_{xx} + c_{12} \varepsilon_{yy} +c_{13} \varepsilon_{zz} - e_{13}
E_z, \quad \sigma_{xy} = (c_{11}- c_{12}) \varepsilon_{xy}/2, \nonumber
\\[10pt]
&& \sigma_{yy} = c_{12} \varepsilon_{xx} + c_{11} \varepsilon_{yy} +c_{13} \varepsilon_{zz} -
e_{31} E_z, \quad  \sigma_{yz}= c_{44} \varepsilon_{yz} - e_{15} E_y, \nonumber
\\[10pt]
&& \sigma_{zz} = c_{13} (\varepsilon_{xx} + \varepsilon_{yy}) +c_{33} \varepsilon_{zz} -
e_{33} E_z, \quad  \sigma_{zx} = c_{44} \varepsilon_{zx} - e_{15} E_x, \quad \nonumber
\\[10pt]
&& D_x = e_{15} \varepsilon_{zx} + \epsilon_{11} E_x, \quad D_y = e_{15} \varepsilon_{yz} +
\epsilon_{11} E_y, \nonumber
\\[10pt]
&& D_z= e_{31} (\varepsilon_{xx} + \varepsilon_{yy}) + e_{33} \varepsilon_{zz} +
\epsilon_{33} E_z + P_{\rm sp}, \label{eq4}
\end{eqnarray}
where $e_{ij}$ and $\epsilon_{ii}$ are piezoelectric and dielectric coefficients; $P_{\rm
sp}$ is the spontaneous polarization. While for the WZ materials the built-in spontaneous
polarization and the principal components of strain are main contributors to the
piezoelectric effect contributions, in ZB materials it is the shear strain components that
may contribute noticeably to the overall properties. For such materials we have the following
constitutive relationships that couple (\ref{eq3}):
\begin{eqnarray}
&& \sigma_{xx} = c_{11}\varepsilon_{xx} +c_{12}\varepsilon_{yy}+c_{12}\varepsilon_{xx}, \quad
\sigma_{yy}=c_{12}\varepsilon_{xx} +c_{11}\varepsilon_{yy}+c_{12}\varepsilon_{zz}, \nonumber
\\[10pt]
&& \sigma_{zz}=c_{12}\varepsilon_{xx} +c_{12}\varepsilon_{yy}+c_{11}\varepsilon_{zz}, \quad
\sigma_{yz}=4c_{44}\varepsilon_{yz}-e_{14}E_x, \nonumber
\\[10pt]
&& \sigma_{zx}=4c_{44}\varepsilon_{zx}-e_{14}E_y, \quad
\sigma_{xy}=4c_{44}\varepsilon_{xy}-e_{14}E_z, \nonumber
\\[10pt]
&&
D_x=e_{14}\varepsilon_{yz}+\epsilon_{11}E_x, \quad
D_y=e_{14}\varepsilon_{zx}+\epsilon_{22}E_y, \quad
 D_z=e_{14}\varepsilon_{xy}+\epsilon_{33}E_z.
 \label{eq5}
\end{eqnarray}
The issue of coupling via boundary conditions remains largely
untouched in the area of modelling piezoelectric semiconductor
nanostructures, in particular when the wetting layer is taken into
account. The formulation of correct boundary conditions in the
latter case was discussed in \cite{Melnik2004}. Recall that the
model we presented in \cite{Melnik2004} accounted for electron
states with arbitrary kinetic energies in the wetting layer.
Effectively, the model we derived for the quantum dot structure with
wetting layer allowed us to demonstrate several important
observations. In particular, electron states that correspond to a
single quantum dot structure with wetting layer will asymptotically
approach one of the two limiting situations: either "pure" quantum
well states far away from the quantum dot region or zero in the case
of a "pure" quantum dot state. This observation must be used, as
explained in \cite{Melnik2004}, for the formulation of general
boundary conditions for the combined quantum-dot/wetting-layer
structure.

The resulting problem we deal with in this paper is a boundary value
problem that is solved with respect to $(u_1, u_2, u_3, \varphi)$.
From a mechanics point of view, the model is derived from a
variational principle applied to the total potential energy which
includes both deformational energy and piezoelectric field
functionals as described in \cite{Melnik2000,Melnik2004ZPiezo}.
Variational difference schemes developed in \cite{Melnik2004ZPiezo},
as well as the finite element formulation developed for computations
in this paper, follow from such a variational representation. Finite
element methodologies have been previously applied to bandstructure
analysis in \cite{Johnson2003,Melnik2004Z,Voss2006}. However, in
these papers the contribution of piezolectric effect was not
accounted for. All works in this area we are aware of are based so
far on the linear theory of elasticity.

Taking into account piezoelectric effect contributions, in the subsequent sections we will
compare the results for bandstructure calculations and the prediction of
optoelectromechanical properties of nanostructures that are obtained with linear and
nonlinear strain models.

%%%%%%%%%%%%%%%%%%%%%%%%%%%%%%%%%%%%%%%%%%
%%%%%%%%%%%%%%%%%%%%%%%%%%%%%%%%%%%%%%%%%%

\section{Bridging the scales to the quantum effect level and exploiting the analogy with
 coupled models of structural mechanics}

Already today, quantum effects play an important role in many optoelectronic devices and
structures. This trend will persist into the future as device miniaturization continues.
Therefore, ideally the full bandstructure transport description is required. However, at
present it is not possible in practice, in particular at the device/structure level. The
reason is simple:
 transport should be computed with a many-particle Hamiltonian for the carriers and the atomic
structures of the device/structure material. This is a task of enormous computational
complexity, not feasible to complete today. Hence, some simplifications need to be made.

One approach is to account for quantum mechanical (and statistical) effects via the
 Wigner-Boltzmann model. Although such a model also
involves substantial computational difficulties, it allows us to
construct a hierarchy of the macroscopic models in a way similar to
those involving fluid dynamics problems and semiconductor device
theory where the continuity (fluid-like) analogy is used for the
model classification (e.g., \cite{Melnik2000H} and references
therein). At present, most of the quantum corrected macroscopic
models are in their infancy as they are usually not able to
adequately include interactions between the electrons and other
particles. New efforts in this direction are currently being
undertaken by a number of authors (e.g., \cite{Bourgade2006}). Since
the problem we are addressing is a {\em multiscale} problem, a
natural way to approach its solution in the above framework could be
to apply a domain decomposition technique. For example, one can use
the quantum mechanical approach in the regions where quantum
mechanical effects are dominant and use continuum-like (e.g.,
hydrodynamic) models in other regions. However, the issue of
coupling such models, e.g., via an interface condition of a typical
domain decomposition methodology or by using other techniques, is
far from trivial and remains largely open.

In this paper we follow another route. While the application of ab
initio and atomistic methodologies are inheritably problematic from
a computational complexity point of view, we resort to averaging
procedures over atomic scales. This can be achieved by a variety of
procedures, including various empirical tight-binding,
pseudopotential, and ${\bf k} \cdot {\bf p}$ approximations. In what
follows, we focus on the latter approximation as a tool for
averaging over atomic scales. The procedure stems from the original
work by Luttinger-Kohn and is based on the effective mass
approximation and the subsequent development of the ${\bf k} \cdot
{\bf p}$ theory. As with any model, the one we develop here relies
on a set of assumptions some of which may not be always fulfilled.
For example, a typical assumption of the ${\bf k} \cdot {\bf p}$
theory that potentials change slowly on the length scale of the
lattice constant could be questionable
 for Metal-Oxide-Semiconductor Field-Effect (MOSFE) devices, e.g. Nevertheless, in bandstructure calculations of LDSNs, the theory
 provides a remarkably flexible tool. Furthermore, while we do not address this issue in detail in this paper,
 it is worthwhile mentioning  that a recent refined approximation of the Luttinger-Kohn
 Hamiltonian, known as the Burt-Foreman correction,  has been developed and tested (see examples and
  further details in \cite{Lassen2004}). This correction allows us to put the effective mass
  theory formalism related to the behaviour of
the envelope functions across interfaces on a much more rigorous mathematical foundation.

For the benefit of the reader, we recall the main premises of the
${\bf k} \cdot {\bf p}$ approximation. Although a number of
methodologies quoted above (such as tight-binding and
pseudo-potential) can provide us with the global dispersion
relationships over the entire Brillouin zone for the bulk material,
to know the main electronic characteristics of the semiconductor,
such as wave functions, we need only the dispersion relationship
over a small wave vector ${\bf k}$ around the band extrema
\cite{Bastard1988}. Indeed, it is well known that most processes in
semiconductors take place near the top of the valence band and at
the bottom of the conduction band \cite{Davies1998}. The wave
functions, characterizing wave propagation in the box volume $\Omega
= \{(x,y,z): 0 \leq x \leq L_x, 0 \leq y \leq L_y, 0 \leq z \leq
L_z,  \}$, are assumed to be in the form
\begin{eqnarray}
&& \Phi_{lmn} ({\bf r}) = (L_x L_y L_z)^{-1/2} \exp [i(k_x x + k_y y
+k_z z)] = \nonumber
\\
&&
(L_x L_y L_z)^{-1/2} \exp (i {\bf k} \cdot {\bf r}),
\end{eqnarray}
where the allowed values of ${\bf k}$
\begin{eqnarray}
{\bf k} = (2 \pi l/L_x, 2 \pi m/ L_y, 2 \pi n /L_z), \quad l, m, n
=0, \pm 1, \pm2,...
\end{eqnarray}
form a 3D ${\bf k}$ space with the origin denoted as $\Gamma$ point.
 Based on this assumption,
 the application of Bloch's theorem in the context of semiconductor
crystals leads to the following
 (Bloch) representation of the wave function in a crystal
\begin{eqnarray}
\Phi_{n {\bf k}} ({\bf r}) = u_{n {\bf k}} \exp (i {\bf k} \cdot
{\bf r})
\end{eqnarray}
with $u_{n {\bf k}}$ being periodic. Since it is easier to find
approximate solutions for the function $u_{n {\bf k}}$ (assumed to
be slowly varying over a small region of ${\bf k}$ space) than for
$\Phi_{n {\bf k}}$ \cite{Davies1998}, we attempt to write the
Schrodinger equation in terms of $u_{n {\bf k}}$. In doing so, the
derivatives of the momentum operator $\hat{\bf p} = -i \hbar \nabla$
acting on the plane waves are simplified to $\hbar {\bf k}$, leading
to a simplification of the Schrodinger equation for the Bloch
functions of the crystal, $\Phi_{n {\bf k}}$, where the terms
depending on ${\bf k}$ are treated as perturbations away from the
solution at ${\bf k}= {\bf 0}$. The presence of the operator ${\bf
k} \cdot \hat{\bf p}$ in the resulting expression renders the name
of the underlying local methodology. In this way, the standard
parabolic approximations of the bands are refined locally in the
${\bf k} \cdot {\bf p}$ theory. This is important for the conduction
band, but even more so for the valence band due to degenerating
light and heavy holes at the $\Gamma$ point.

It is worthwhile mentioning that another major reason for this
particular choice of averaging lies with the fact that the ${\bf k}
\cdot {\bf p}$ treats the bandstructure in a continuum-like manner,
albeit allowing to incorporate many important effects acting at
different scales which is very important for the problems like ours.
This includes mechanical and electromechanical effects. As strain
and piezoeffect
 are key contributors to changes in optoelectromechanical properties, the models described in
the previous section must be incorporated in the bandstructure
calculation. In order to do that, we first remind the reader that
the accuracy of approximations based on the ${\bf k} \cdot {\bf p}$
theory depends on the functional space where the envelope function
is considered. In fact, we put subbands within conduction and
valence bands of the semiconductor material into correspondence to
the basis functions that span such a space. The number of such
functions varies in applications, but typically ranges from 1 to 8
to ensure computational feasibility of the problem. For WZ
materials, 8 functions should usually be included due to spin-orbit,
crystal-field splitting, as well as conduction/valence band mixing
effects. This corresponds to 6 valence subbands and 2 conduction
subbands that account for spin up and spin down situations. Given
that the Hamiltonian in the ${\bf k} \cdot {\bf p}$ theory can be
represented as
\begin{eqnarray}
H = - \frac{\hbar^2}{2 m_0} \nabla_i {\mathcal H}_{ij}^{(m, n)}
({\bf r}) \nabla_j, \label{eq6}
\end{eqnarray}
the problem at hand can be formulated as an eigenvalue partial differential equation problem
\begin{eqnarray}
H \Psi = E \Psi. \label{eq7}
\end{eqnarray}
The standard Kohn-Luttinger representation of the Hamiltonian in the
form of (\ref{eq6}) is well documented in the literature and we
refer the reader interested in details to books by G. Bastard and J.
Singh, e.g. \cite{Bastard1988,Singh1993}.  The problem (\ref{eq7})
should be solved with respect to eigenpair $(\Psi, E)$, where $E$ is
the electron/hole energy and $\Psi$ is the wave vector with
dimensionality of the functional space chosen. For example, in the
situation discussed above we have
\begin{eqnarray}
\Psi = (\psi_S^{\uparrow}, \psi_X^{\uparrow}, \psi_Y^{\uparrow}, \psi_Z^{\uparrow},
\psi_S^{\downarrow}, \psi_S^{\downarrow}, \psi_S^{\downarrow}, \psi_S^{\downarrow})^T,
\label{eq8}
\end{eqnarray}
where the subindex S denotes the wave function component of the
conduction band and $\psi_X^{\uparrow} \equiv (|X>| \uparrow)$ is
the wave function component that corresponds to the X Bloch function
of the valence band when the spin function of the missing electron
is up. In the Hamiltonian representation (\ref{eq6})  $\mathcal H$
is the energy functional, defined either by the standard
Kohn-Luttinger Hamiltonian (as mentioned above) or by the
Burt-Foreman modification \cite{Lassen2004}. It represents the
kinetic energy plus a nonuniform potential field $V$ and other
effects contributing to the total potential energy of the system, as
specified further in Section 4. Other notations are standard:
$\hbar$ is the Planck constant, $m_0$ is the free electron mass,
${\bf r}= (x_1 \equiv x, x_2 \equiv y, x_3 \equiv z)$, while the
superindeces $(m, n)$ are used to denote
 the basis of the space for the wave function, which in the case discussed above (6 valance and 2 conduction
subbands) would lead to an $8X8$ Hamiltonian.

Four equations (\ref{eq3}) of coupled electromechanics and the
eigenvalue PDE problem (\ref{eq7}) constitute a coupled
Schrodinger-Poisson model that provides a general framework for
incorporating nonlinear effects. Similar to the top-down approach,
where coupling procedures are well established, in the bottom-up
approach applied here the coupling between Schrodinger and Poisson
equations is essential, in particular at the level of quantum device
modelling.

\section{Coupling strain with bandstructure calculations in the variational formulation}

Since the fundamental work \cite{Bir1974}, strain effects firmly took their place in the
models for bandstructure calculations of semiconductor structures. Many important examples of
such calculations based on strained Rasba-Sheka-Pikus Hamiltonian have recently been provided
in the context of low-dimensional semiconductor nanostructures, including quantum dots (e.g.,
\cite{Fonoberov2003}). However, all current models we are aware of have been based on linear
theories. The first question to ask is whether material nonlinearities, expressed by
stress-strain relationships, may become important for such calculations. The usual argument
for using the linear stress-strain relationships in this field of applications is based on
the fact that strain is indeed of order of magnitudes smaller of the elastic limits. Although
 this argument may fail at the device level modelling, it remains valid for the
structures of interest in the present paper. At a more fundamental
level, however, elastic and dielectric coefficients may become
 nonlinear due to geometry of the structure which will lead to nonlinear
  stress-strain relationships. To address this issue
rigorously, one has to look critically at the standard Keating model
where the parameters of the model are unit-cell dimension dependent.
In the general case, we may need to account for many body
interactions of higher order to reflect asymmetry of the interatomic
potential. This issue is outside of the scope of the present paper.
Instead, we will focus on a consequence of this issue that leads to
geometric nonlinearities. Indeed, one of the major drawbacks of the
current models for bandstructure calculations is that they are not
able to resolve adequately strain nonhomogeneities due to the
application of the original representation of \cite{Bir1974}
 based the infinitesimal theory with Cauchy relationships between strain and displacements.
 As we demonstrate in the next section, this approximation is inadequate when we have to deal
 with geometric irregularities of low-dimensional semiconductor structures. In the latter
 case, by using the variation of deformation $\delta \varepsilon_{ij}$ which is induced by the
 variation in displacements $\delta u_i$, we describe the new position $\xi_i = x_i+u_i$ of the
 material particle (with initial coordinates $x_i$) after deformation. This leads naturally to the
 formulation of the problem where the general nonlinear Green-Lagrange relationship for strain
  can be accounted for. As we pointed out, the variational formulation of the problem at
  hand has been used before in the context of finite element implementations (e.g.,
  \cite{Johnson2003,Melnik2004Z,Voss2006}). Two new features that have been developed
  in this paper include (a) consistent coupled treatment of the piezoelectric effect
  and (b) the ability to incorporate geometric nonlinearities into the model. As we demonstrate
   in the next section, the influence of these features on
the bandstructure calculations, and therefore optoelectromechanical properties of the
modelled structures, could be substantial.

The case of weakly coupled piezoelectricity was analyzed previously in \cite{Melnik1998}
where convergence results in Sobolev classes of generalized solutions were established.
Indeed, in some special cases, in particular for simple boundary conditions, the problem can
be addressed semi-analytically. In the context of quantum dots, this has been done in
\cite{Pan2002} by using Green's function approach. Strongly coupled problems of
piezoelectricity have been analyzed previously from variational scheme perspectives in
\cite{Melnik2004ZPiezo}. In the latter case, the steady-state formulation given by
(\ref{eq3}) should be understood in a variational sense as:
\begin{eqnarray}
\int_{\tilde{V}} \left[ - \sigma^T (\delta \varepsilon_L + \delta \varepsilon_N) + D \delta E
\right ] d v =0, \label{eq9}
\end{eqnarray}
where both piezoelectric stress and the nonlinear part of strain are taken into account via
constitutive relationships with the total variation of deformation given by its linear and
nonlinear parts $\delta \varepsilon = \delta \varepsilon_L + \delta \varepsilon_N$.

The above problem is coupled to the eigenvalue PDE problem, also understood in a weak sense.
That is we seek the solution to the following problem
\begin{eqnarray}
\Phi(\Psi) \rightarrow \min, \quad \Psi \equiv - \frac{\hbar^2}{2 m_0} \int_{\tilde{V}}
(\nabla \Psi)^T {\mathcal H}^{(m, n)} \nabla \Psi d v - E \int_{\tilde{V}} \Psi^T \Psi d v,
\label{eq10}
\end{eqnarray}
where the Hamiltonian is given by (\ref{eq6}).  As usual in the
${\bf k} \cdot {\bf p}$ theory, formal representation can be reduced
to the sum of constant and ${\bf k}$-dependent energies ($H=\bar{H}
+ V$):
\begin{eqnarray}
\bar{H}= H_0 + \sum_{i=1}^3 H_i, \label{eq11}
\end{eqnarray}
where $H_0$, derived from the standard Kane Hamiltonian at ${\bf
k}={\bf 0}$ (e.g., \cite{Bastard1988}), accounts for the
spin-splitting effects; $H_1$ is the contribution due to the kinetic
part of the microscopic Hamiltonian unit cell averaged by the
respective Bloch functions, S, X, Y, or Z; $H_2$ is the
strain-dependent part of the Hamiltonian, and $H_3$ is the energy of
unstrained conduction/valence band edges.

In what follows we consider an example of modelling nanostructures
based on the model described above. The main emphasis is given to a
pyramidal quantum dot residing on a wetting layer. In
\cite{Melnik2004} it was demonstrated that the electronic states in
the wetting layer may influence electronic states in the dot and
vice versa and, hence, the thin layer on which the quantum dot
resides cannot be excluded from the computational domain as it is
often done in the literature.  This brings additional difficulties
in the computational implementation of this multiscale problem.

%%%%%%%%%%%%%%%%%%%%%%
\section{Computational experiments on quantum dot nanostructures}

Our representative example in this section concerns InAs/GaAs quantum dot structures of
pyramidal shape. Such self-assembled structures are grown experimentally via
Stranski-Krastanov methodology and have been studied before with simplified models of linear
elasticity (e.g., \cite{Grundmann1995,Kuo2006}). As we have already mentioned, the present
study has a number of new features. Firstly, we take the full electromechanical coupling as
well as the wetting layer into account. Secondly, we analyze the influence of nonlinear
strain components in modelling such quantum dot structures.

%% AlN - piezoelectric constant is big...

\begin{figure}
\begin{center}
{\epsfig{file=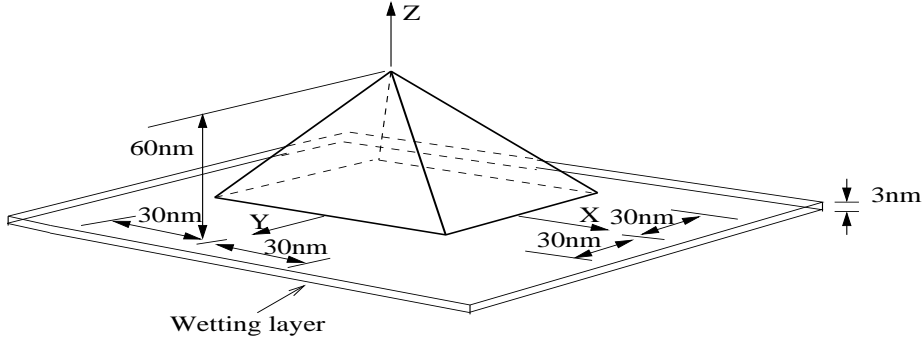, height=4.5cm, width=12.2cm}}
%\end{tabular}
%\vspace{-0.8cm}
\caption{Geometric dimensions of the representative pyramidal quantum dot.}
\label{figure1}
\end{center}
%\vspace{-0.1cm}
\end{figure}
%\vspace{-1cm}

\begin{figure}
\begin{center}
{\epsfig{file=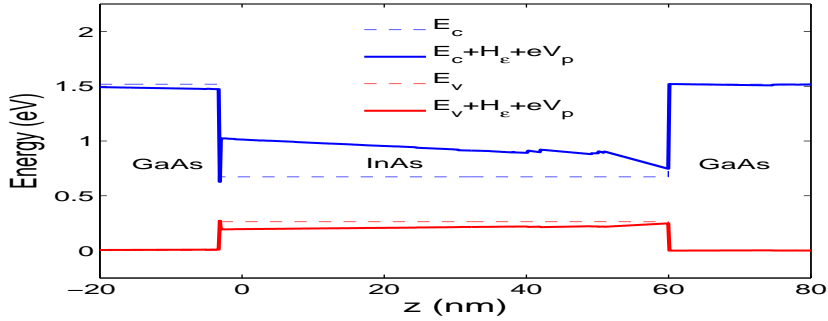, height=4.5cm, width=12.2cm}}
%\end{tabular}
%\vspace{-0.8cm}
\caption{Energy levels of the conduction and valence bands under the linear strain
approximation.} \label{figure2}
\end{center}
%\vspace{-0.1cm}
\end{figure}

All our computational examples below are for the quantum dot structure presented in Fig.
\ref{figure1} where geometric dimensions are also given. Pyramidal shapes of quantum dot
structures, such as the one we analyze, have been confirmed by experimental techniques,
including high resolution electron microscopy. The entire structure, consisting of the InAs
dot sitting on the wetting layer, is embedded in a (spherical) GaAs matrix. In such
semiconductor materials one expects changes in the optoelectromechanical properties due to
strain effects. However, up until now, such changes have been quantified with linear theories
only. Hence, as the first step, in Fig. \ref{figure2} we present energy levels for both
conduction and valence bands by using the conventional methodology based on the linear
approximation. This result is given at the center of the dot ($x=0,y=0$) along vertical
z-axis.

%\vspace{-1cm}

A straightforward  generalization of the linear theory in our context is to account for
 the large deformation gradient $\nabla u_3$. This gradient is responsible for
 the dominant nonlinear strain effect due to the lattice mismatch in the growth
 direction. While in the x and y
 directions the strain components will be identical in this case to the von Karman type model,
in the z-direction they differ. As all the shear strain components are assumed in this case
[Case I] to be linear, the Green-Lagrange strain components in this model take the form:
\begin{eqnarray}
&& \varepsilon_{xx}=\frac{\partial u_1}{\partial x} +\frac{1}{2}\left(\frac{\partial
u_3}{\partial x}\right)^2 \;, \quad \varepsilon_{yy}=\frac{\partial u_2}{\partial y}
+\frac{1}{2}\left(\frac{\partial u_3}{\partial y}\right)^2 \;, \nonumber
\\[10pt]
&& \varepsilon_{zz}=\frac{\partial u_3}{\partial z} +\frac{1}{2}\left(\frac{\partial
u_3}{\partial z}\right)^2 \;. \label{eq12}
\end{eqnarray}

\begin{figure}
\begin{center}
{\epsfig{file=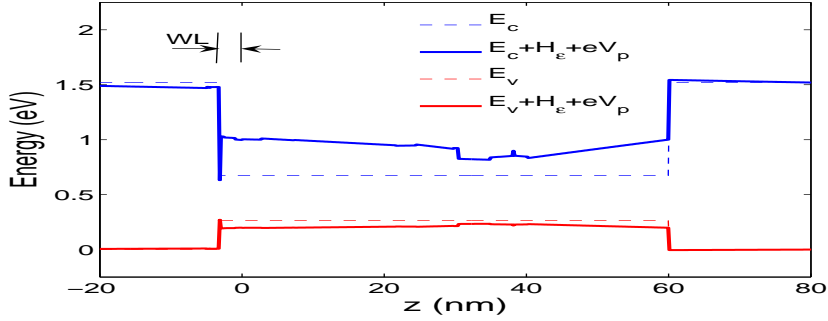, height=4.5cm, width=12.2cm}}
%\end{tabular}
%\vspace{-0.8cm}
\caption{Energy levels of the conduction and valence bands accounting for the large
deformation gradient in the growth direction.} \label{figure3}
\end{center}
%\vspace{-0.1cm}
\end{figure}

In this case, closer to the base of the dot, the behaviour of the energy levels are the same
as in the linear case. However, the situation changes at the tip of the dot as can be seen
from Fig. \ref{figure3}. As before, the result is presented at the center of the dot
($x=0,y=0$) along vertical z-axis.

%\vspace{-1cm}

\begin{figure}
\begin{center}
{\epsfig{file=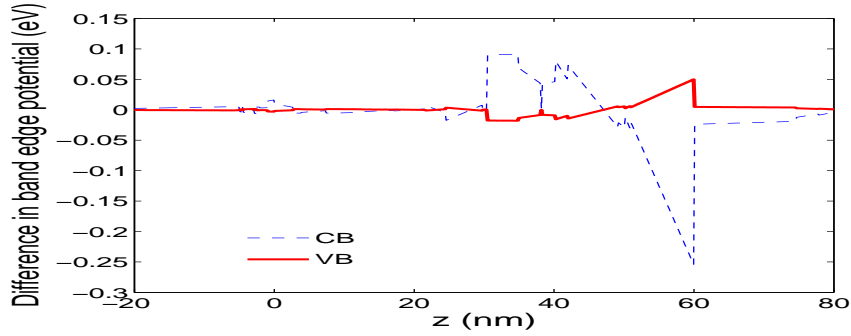, height=4.5cm, width=12.2cm}}
%\end{tabular}
%\vspace{-0.8cm}
\caption{Quantifying nonlinear contributions in the band edge potentials, accounting  the
large deformation gradient in the growth direction.} \label{figure4}
\end{center}
%\vspace{-0.1cm}
\end{figure}

Next, we have analyzed and quantified the difference between this nonlinear case and the
linear case, conventionally used in these calculations. In Fig. \ref{figure4} we present the
difference in the band edge potential for the conduction and valence bands. This result
demonstrates that nonlinear strain contributions could be substantial in calculating energy
levels, and hence in predicting optoelectronic properties of nanostructures. In our case
these contributions are particular pronounced for the conduction band.

%\vspace{-1cm}

The comparisons indicate the maximum deviation from the linear theory near the top vertex of
the pyramid of $60 {\rm nm}$ high, not at the wetting layer. Indeed, due to the lattice
mismatch at the wetting layer and pyramid interface, components like ${(\partial u_1/\partial
x)}^2$ and ${(\partial u_2/ \partial y)}^2$ will become important, rather than ${(\partial
u_3/
\partial x)}^2$ or ${(\partial u_3/ \partial y)}^2$ which are due mainly to the Poisson effect
in this problem.

\begin{figure}
\begin{center}
{\epsfig{file=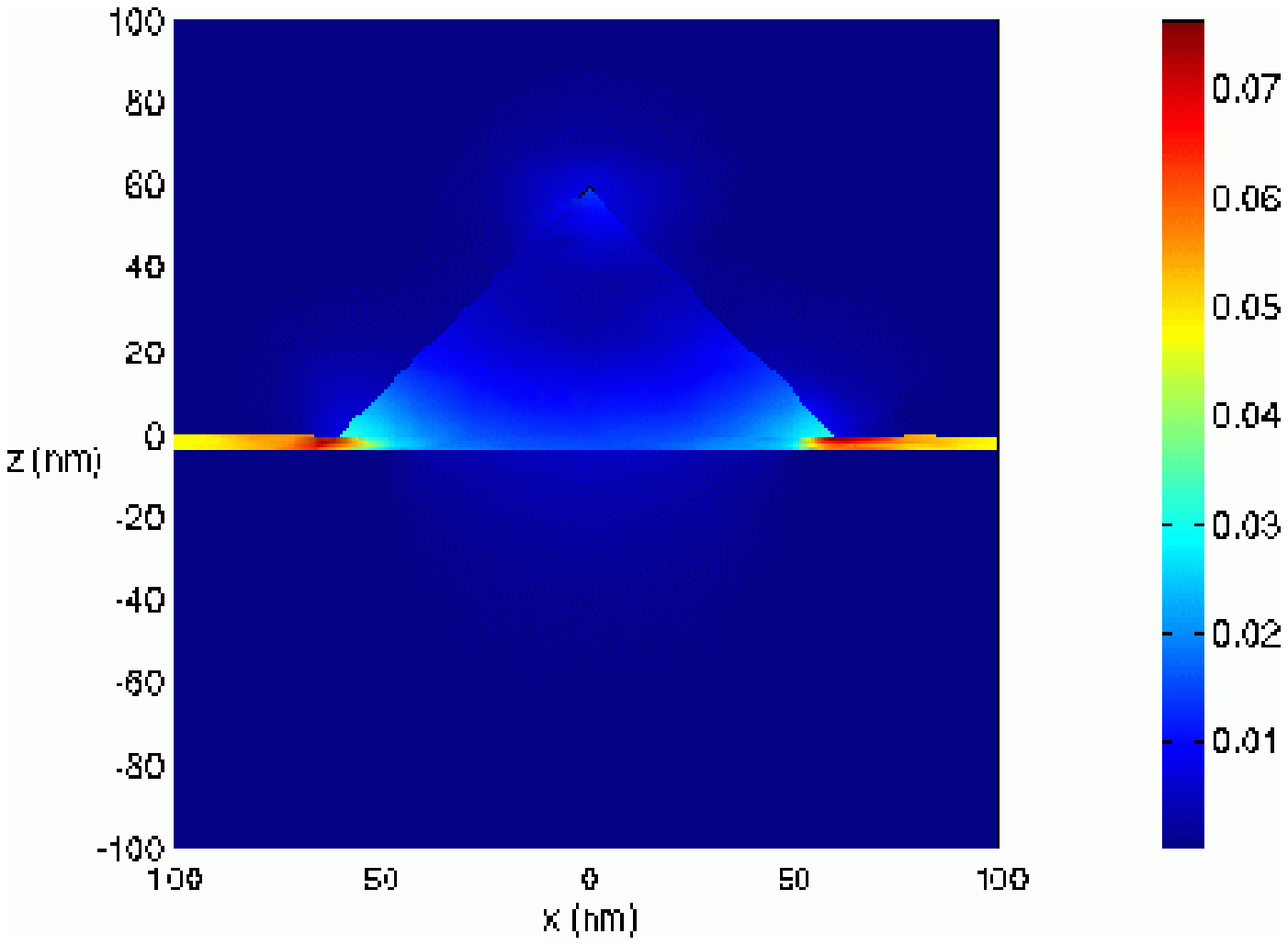, height=4cm, width=6.1cm}}
{\epsfig{file=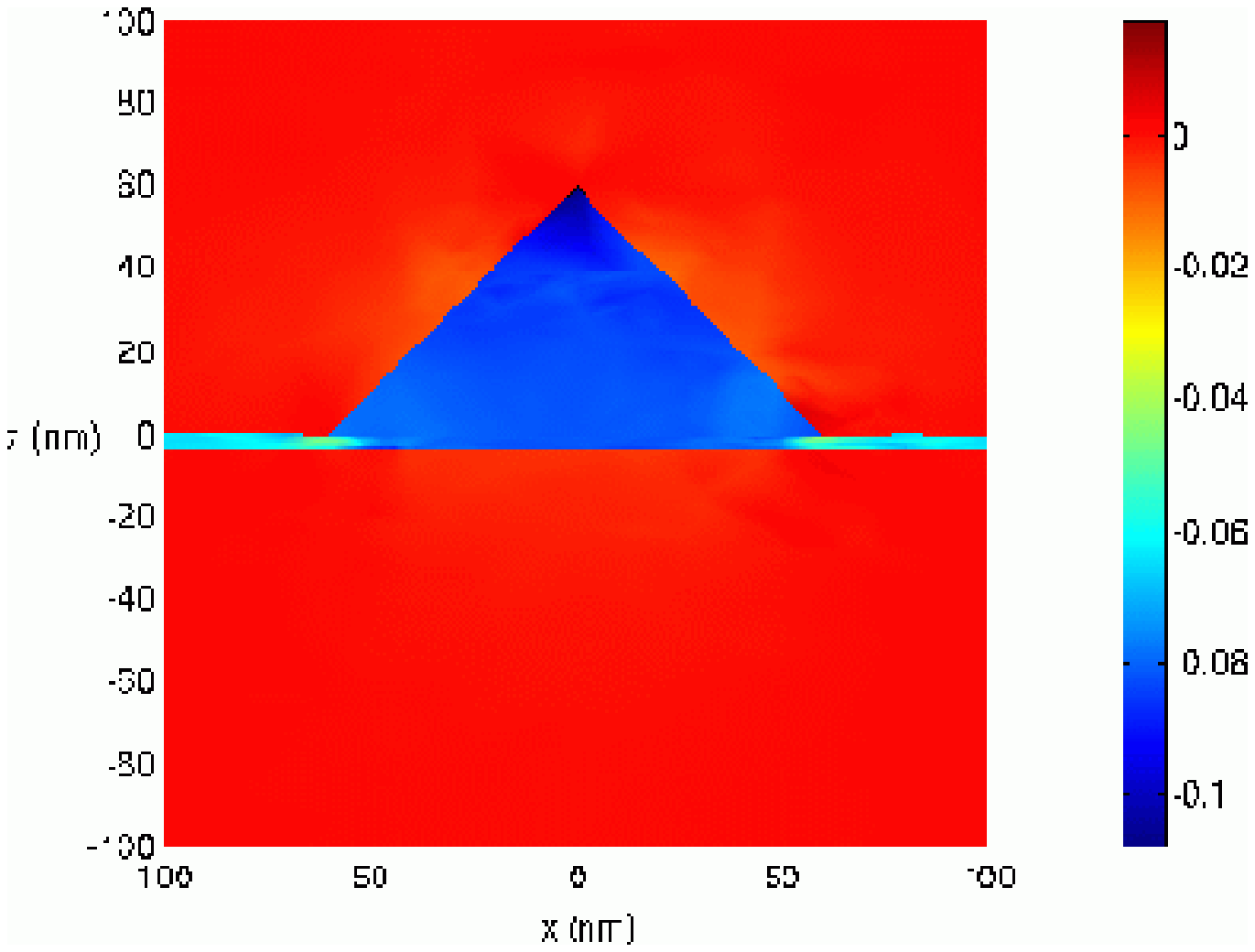, height=4cm, width=6.1cm}}
%\end{tabular}
%\vspace{-0.8cm}
\caption{Biaxial and isotropic strains at y=0.}
\label{figure5}
\end{center}
%\vspace{-0.1cm}
\end{figure}

These differences between the results produced with the linear and nonlinear models are hard
to quantify without the band edge potentials, as presented above. Indeed, we have calculated
a number of other characteristics, including biaxial strain
\begin{eqnarray}
\varepsilon_b={(\varepsilon_{11}- \varepsilon_{22})}^2+{(\varepsilon_{22}-
\varepsilon_{33})}^2 +{(\varepsilon_{33}- \varepsilon_{11})}^2 \label{eq13}
\end{eqnarray}
 and isotropic strain
\begin{eqnarray}
 \varepsilon_i= \varepsilon_{11}+
\varepsilon_{22}+ \varepsilon_{33}. \label{eq14}
\end{eqnarray}
 In Fig. \ref{figure5} we present biaxial and isotropic strains
in the y=0 plane.

%\vspace{-1cm}

\begin{figure}
\begin{center}
{\epsfig{file=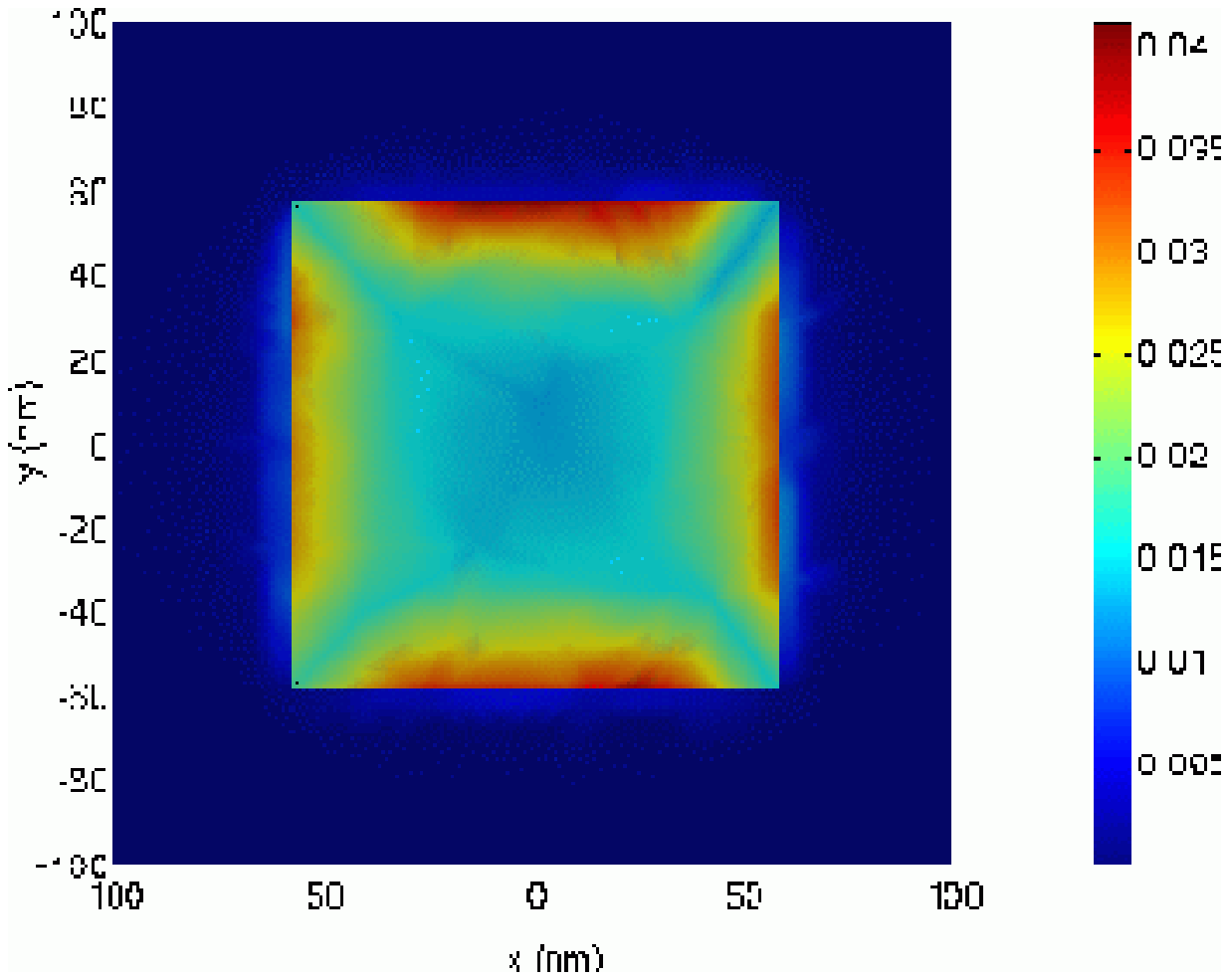, height=4cm, width=6.1cm}}
{\epsfig{file=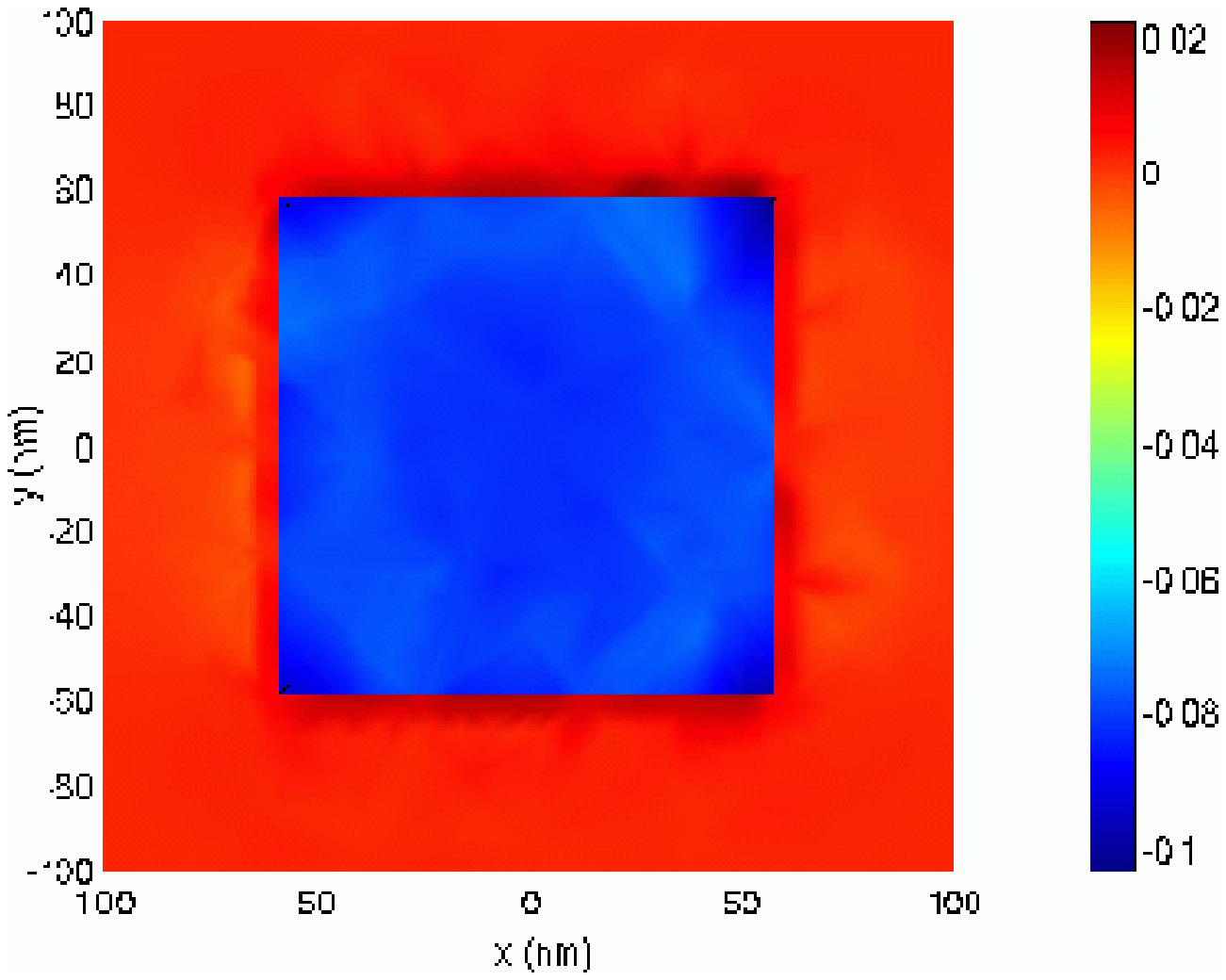, height=4cm, width=6.1cm}}
%\end{tabular}
%\vspace{-0.8cm}
\caption{Biaxial and isotropic strains at z=1.5nm}
\label{figure6}
\end{center}
%\vspace{-0.1cm}
\end{figure}

These characteristics are shown also just above the wetting layer, at z=1.5nm, in Fig.
\ref{figure6}. In both of the above situations we chose to present the results of
calculations produced with the conventional methodology, while noting that calculations
obtained with nonlinear contributions look very similar.

%\vspace{-1cm}

\begin{figure}
\begin{center}
{\epsfig{file=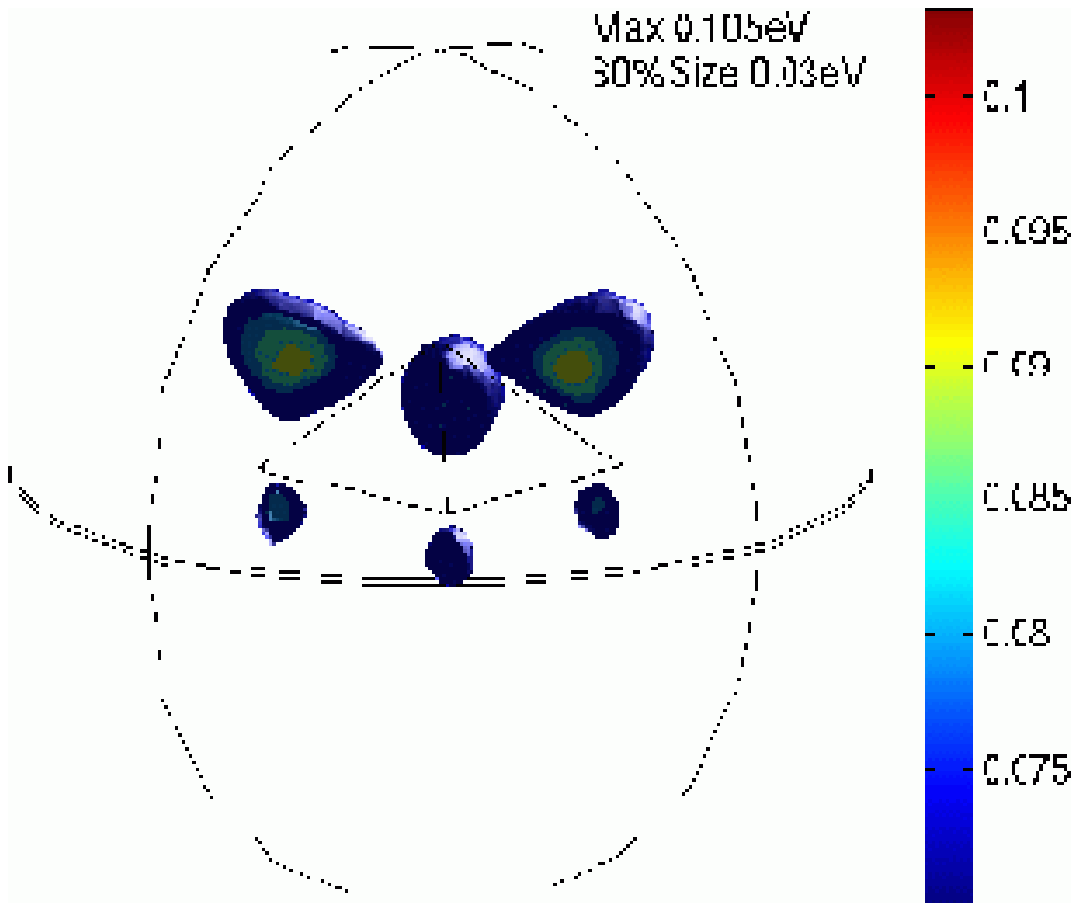, height=4.5cm, width=6.1cm}}
{\epsfig{file=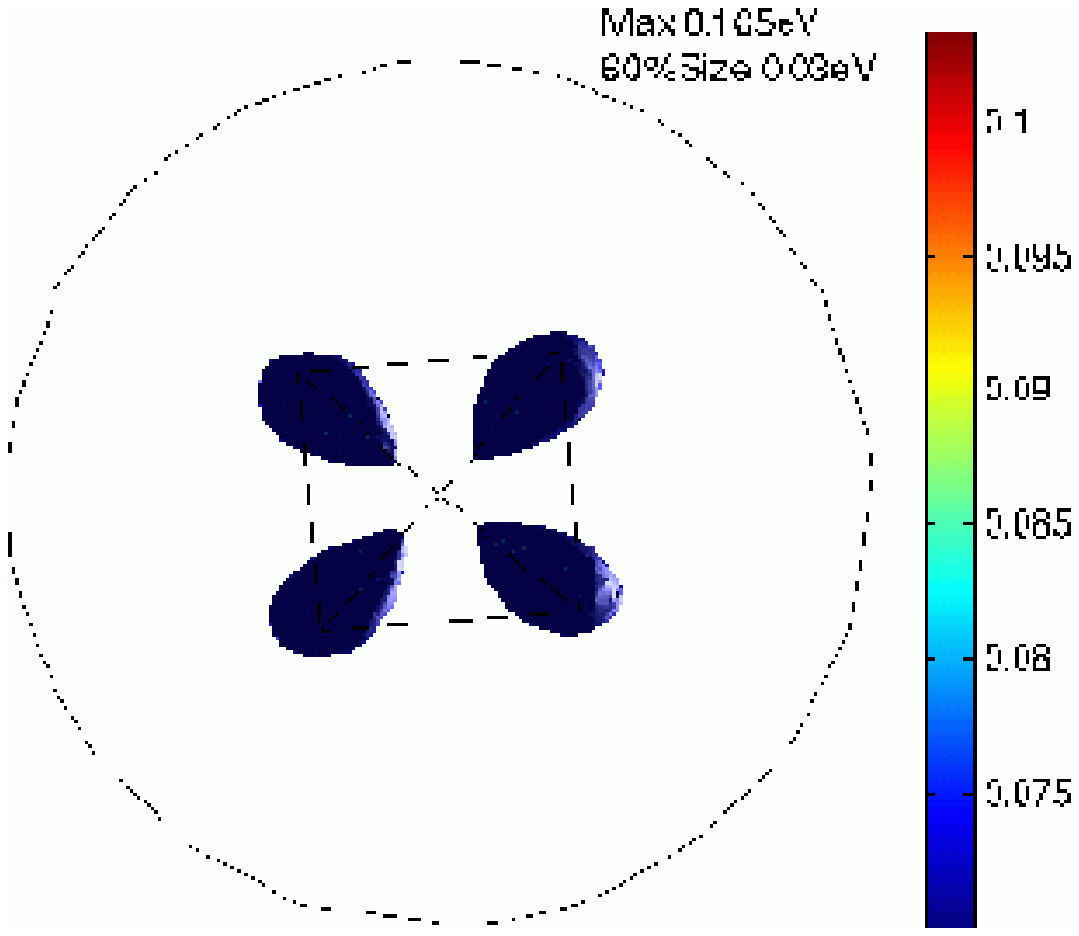,height=4.5cm, width=6.1cm}}
%\end{tabular}
%\vspace{-0.8cm}
\caption{Distribution of the piezoelectric potential in the quantum dot.}
\label{figure7}
\end{center}
%\vspace{-0.1cm}
\end{figure}

    In Fig. \ref{figure7} we present
the distribution of the piezoelectric potential in the quantum dot under consideration (as
well as its projection from the top). Recall that results with uncoupled or semi-coupled
models usually demonstrate maxima of the piezopotential outside of ZB quantum dot structures
only, in particular when the dot is truncated. This is, of course, not the case for the
coupled model applied in the current situation. The local extrema of the piezoelectric
potential near the QD top can be well reproduced, as demonstrated by Fig. \ref{figure7}. As
expected, the distribution is symmetric at the bottom and at the top of the pyramid. We note
also that predicting optoelectromechanical properties of hexagonal (WZ) materials usually
requires to account for an additional effect of spontaneous polarization, typically
negligible in ZB materials.  As described in Section 2, our model is capable of dealing with
both types of the materials.

%\vspace{-1cm}

Next, we have calculated eigenstates of the structure. The ground state, presented in Fig. 8,
has been calculated accounting for strain and piezoelectric effects. As seen from the figure,
the state is fully confined.

\begin{figure}
\begin{center}
{\epsfig{file=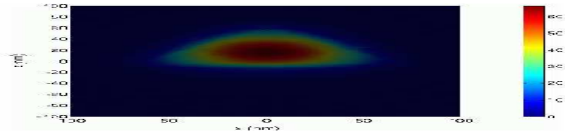, height=4.5cm, width=12.2cm}}
%\end{tabular}
%\vspace{-0.8cm}
\caption{Ground state of the quantum dot nanostructure.} \label{figure8}
\end{center}
%\vspace{-0.1cm}
\end{figure}
%\vspace{-1cm}

In Fig. 9 the next four eigenstates of the nanostructure are presented. As confinement
visualization looks similar for both linear and nonlinear cases, we shall provide calculated
numerical values that demonstrate the influence of piezoelectric effect as well as nonlinear
contributions on electronic states.

\begin{figure}
\begin{center}
{\epsfig{file=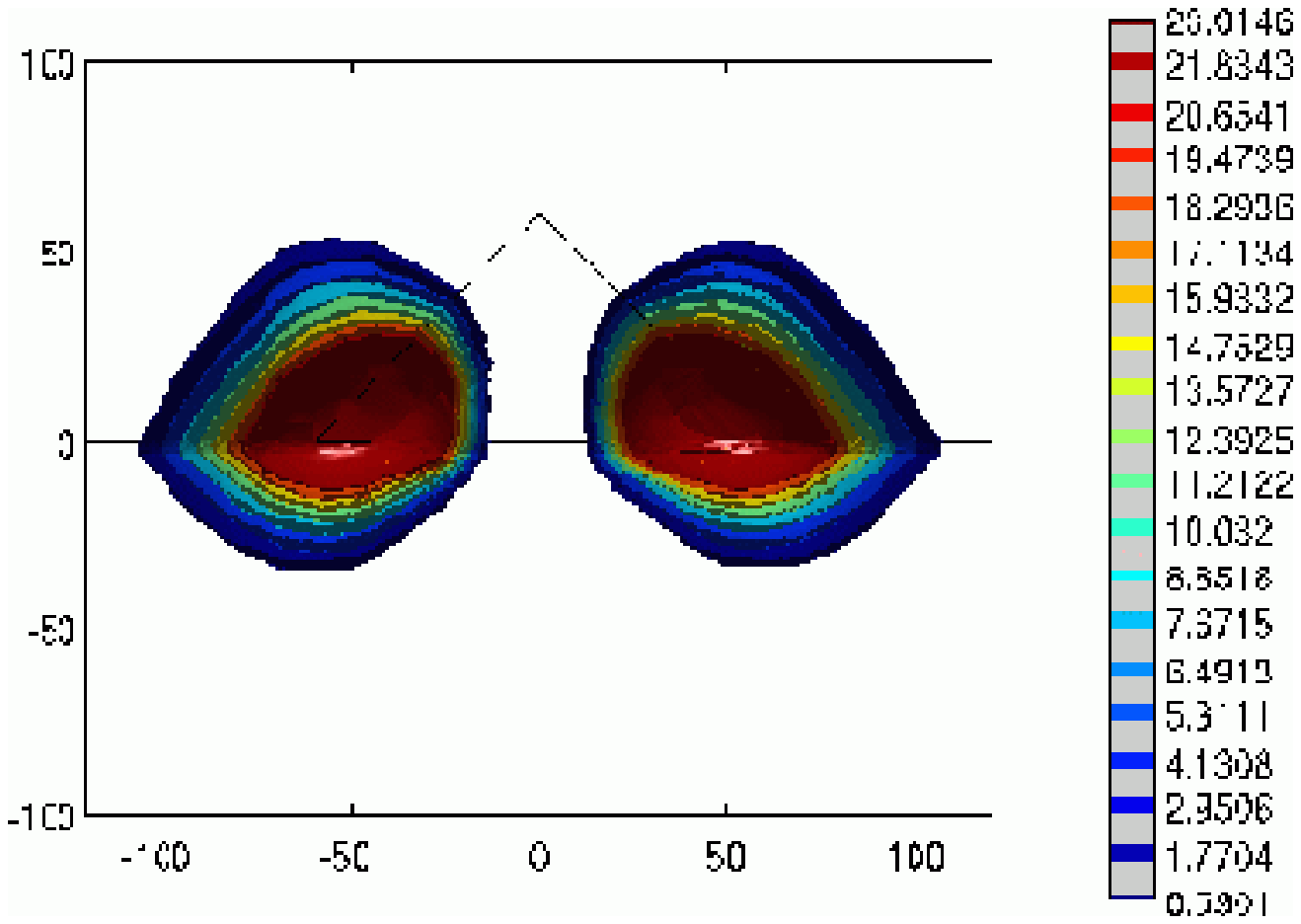, height=4cm, width=6.1cm}}
{\epsfig{file=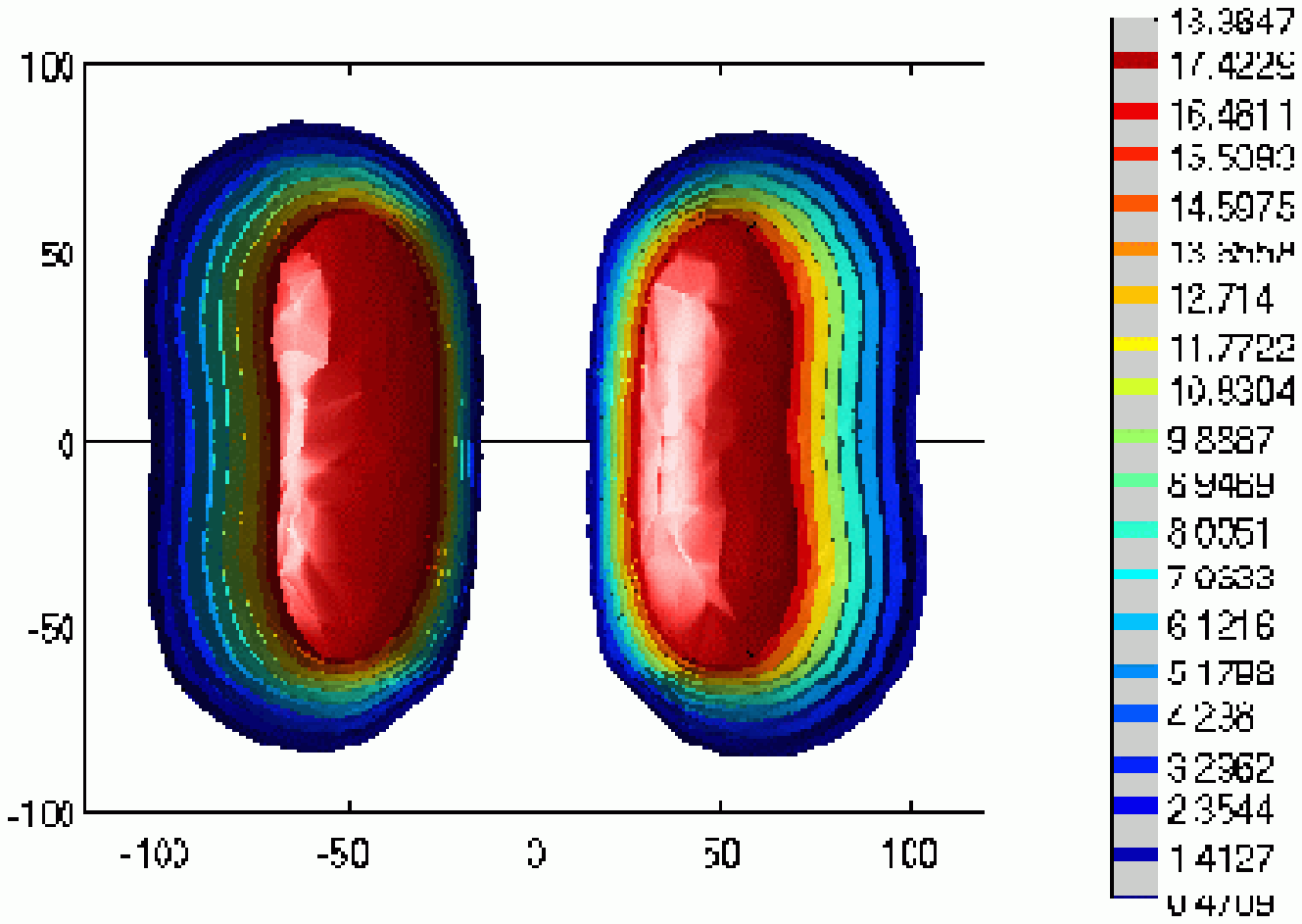, height=4cm, width=6.1cm}} \\
{\epsfig{file=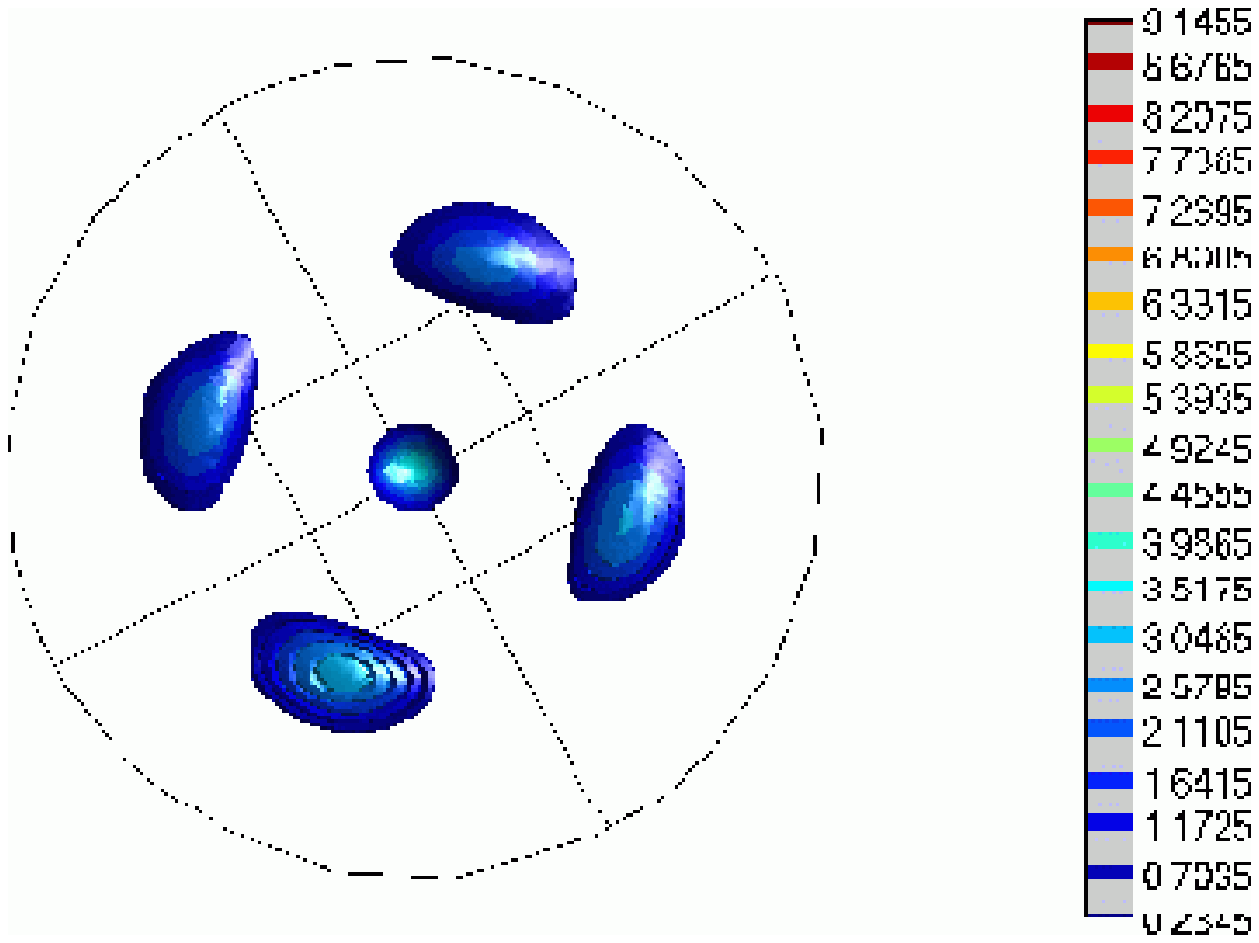, height=4cm, width=6.1cm}}
{\epsfig{file=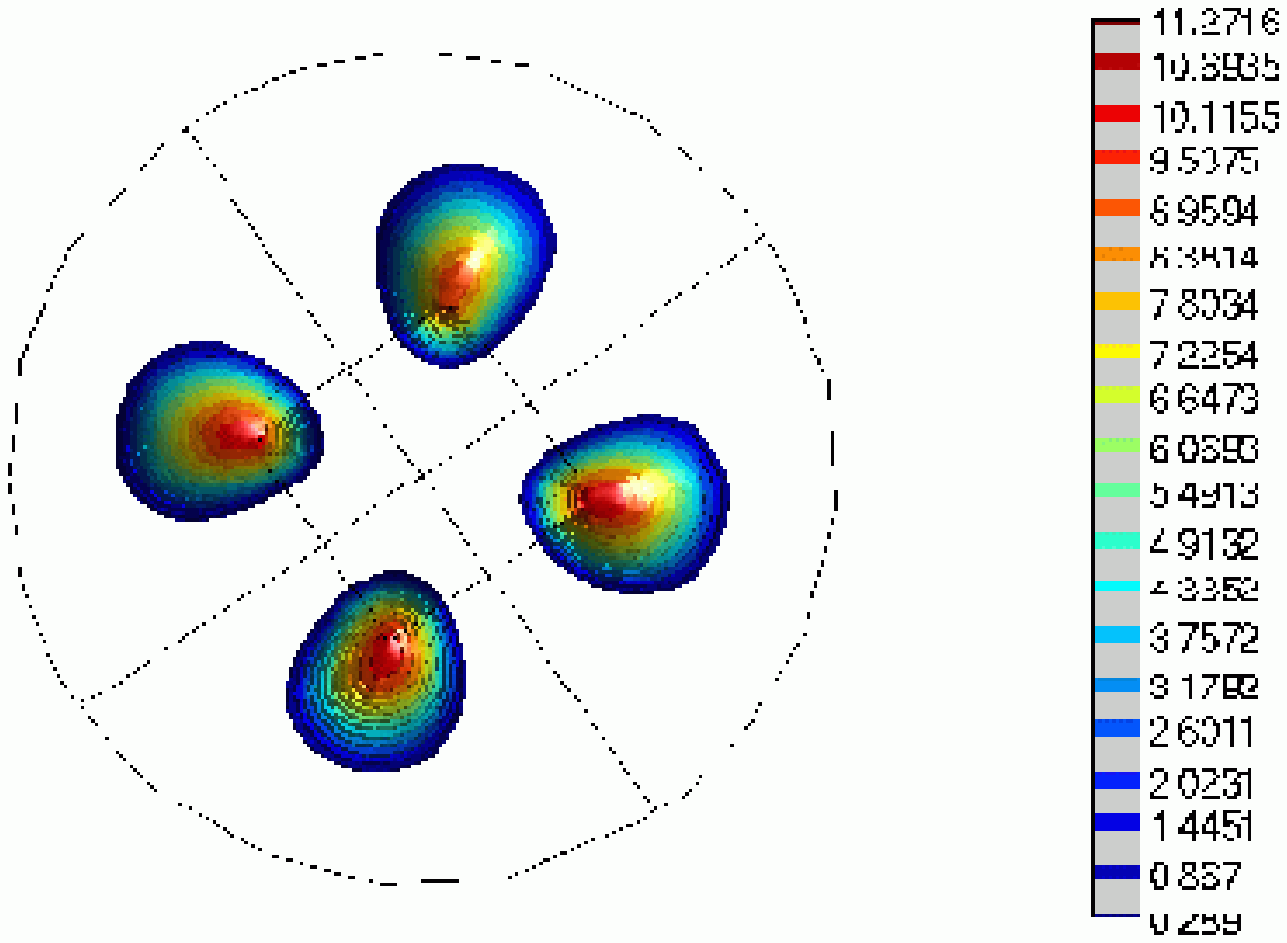, height=4cm, width=6.1cm}}
%\end{tabular}
%\vspace{-0.8cm}
\caption{Electronic confinement in the quantum dot nanostructure: the 2nd, 3rd, 4th, and 5th
eigenstates.}
\end{center}
%\vspace{-0.1cm}
\end{figure}
%\vspace{-1cm}

Before doing that, we recall that the only nonlinear model we considered so far was the model
accounting for the large deformation gradient in the growth direction. We have also analyzed
the case where we account for the large deformation gradients $\nabla u_1$ and $\nabla u_2$,
 responsible for the dominant nonlinear strain effects due to the lattice mismatch in the $x-y$
 plane, normal to
the growth direction. In this case [Case II] the Green-Lagrange strain components are:
\begin{eqnarray}
\varepsilon_{xx}=\frac{\partial u_1}{\partial x} +\frac{1}{2}\left(\frac{\partial
u_1}{\partial x}\right)^2 +\frac{1}{2}\left(\frac{\partial u_2}{\partial x}\right)^2 \;,
\nonumber
\\[10pt]
 \varepsilon_{yy}=\frac{\partial u_2}{\partial y} +\frac{1}{2}\left(\frac{\partial
u_1}{\partial y}\right)^2 +\frac{1}{2}\left(\frac{\partial u_2}{\partial y}\right)^2 \;,
\nonumber
\\[10pt]
\varepsilon_{zz}=\frac{\partial u_3}{\partial z} +\frac{1}{2}\left(\frac{\partial
u_1}{\partial z}\right)^2 +\frac{1}{2}\left(\frac{\partial u_2}{\partial z}\right)^2 \;,
\label{eq15}
\end{eqnarray}
while all the shear strain components remain linear. As expected in this case, the results
 show nonlinear effect contributions due to the nonlinear relaxation of the interfacial strain
between the InAs wetting layer and the GaAs matrix.

Next, we implemented the full nonlinear strain model [Case III] in
the developed finite element code, based  on the general
Green-Lagrange strain
\begin{eqnarray}
\mbox{\boldmath$\varepsilon$} = \frac{1}{2}({\bf F}^T {\bf F} - {\bf I}) \;, \label{eq16}
\end{eqnarray}
where ${\bf F}$ is the deformation gradient and ${\bf I}$ is the identity matrix.  In Figs.
10--12 the results are presented along the z-axis (y=0) for the three cases: x=0, 30, and 60
${\rm nm}$, respectively. Although differences with calculations obtained with the linear
model are clearly observed, we quantify them in Table 1. The first column in the table is the
number of the corresponding eigenstate.  The second column gives eigenstates obtained with
the linear theory, accounting for strain only; the third column gives eigenstates obtained
with the linear theory, accounting for strain and piezo effects. The remaining columns give
the results obtained with the nonlinear models based on Cases I, II, and III, respectively,
as described above. All values are given in eV.

In all finite element computations reported here we applied
tetrahedral elements with quadratic Lagrangian interpolation
function. The global convergence was analyzed by refining the finite
element mesh and ensuring that the difference between the last two
subsequent refinements is negligible in $L_2$. For the solution of
the discretized equations we used GMRES with incomplete Cholesky
factorization. The maximum number of iteration was set to 500 and
the $L_2$ error tolerance was set as $1 \times 10^{-6}$. The pyramid
was embedded into a sphere. The loading conditions were simulated by
subjecting the outer surface of the sphere to Dirichlet boundary
conditions in displacement. The pyramid structure was allowed to
equilibrate under the combined effect of lattice misfit induced
strain and piezoelectric polarization.

\begin{figure}
\begin{center}
{\epsfig{file=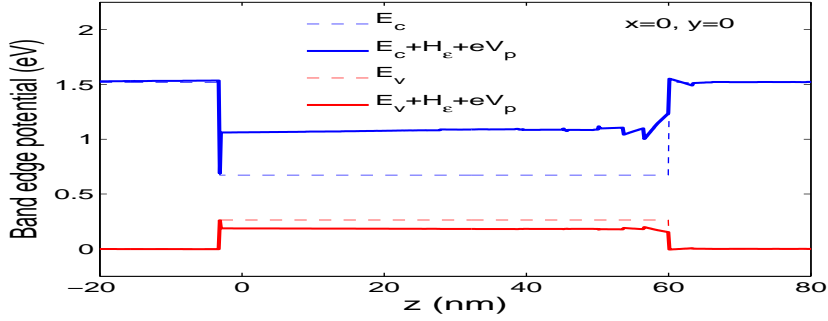, height=4.5cm, width=12.2cm}}
%\end{tabular}
%\vspace{-0.8cm}
\caption{The band edge potential with the full nonlinear strain model (x=0, y=0).}
\end{center}
%\vspace{-0.1cm}
\end{figure}
%\vspace{-1cm}

\begin{figure}
\begin{center}
{\epsfig{file=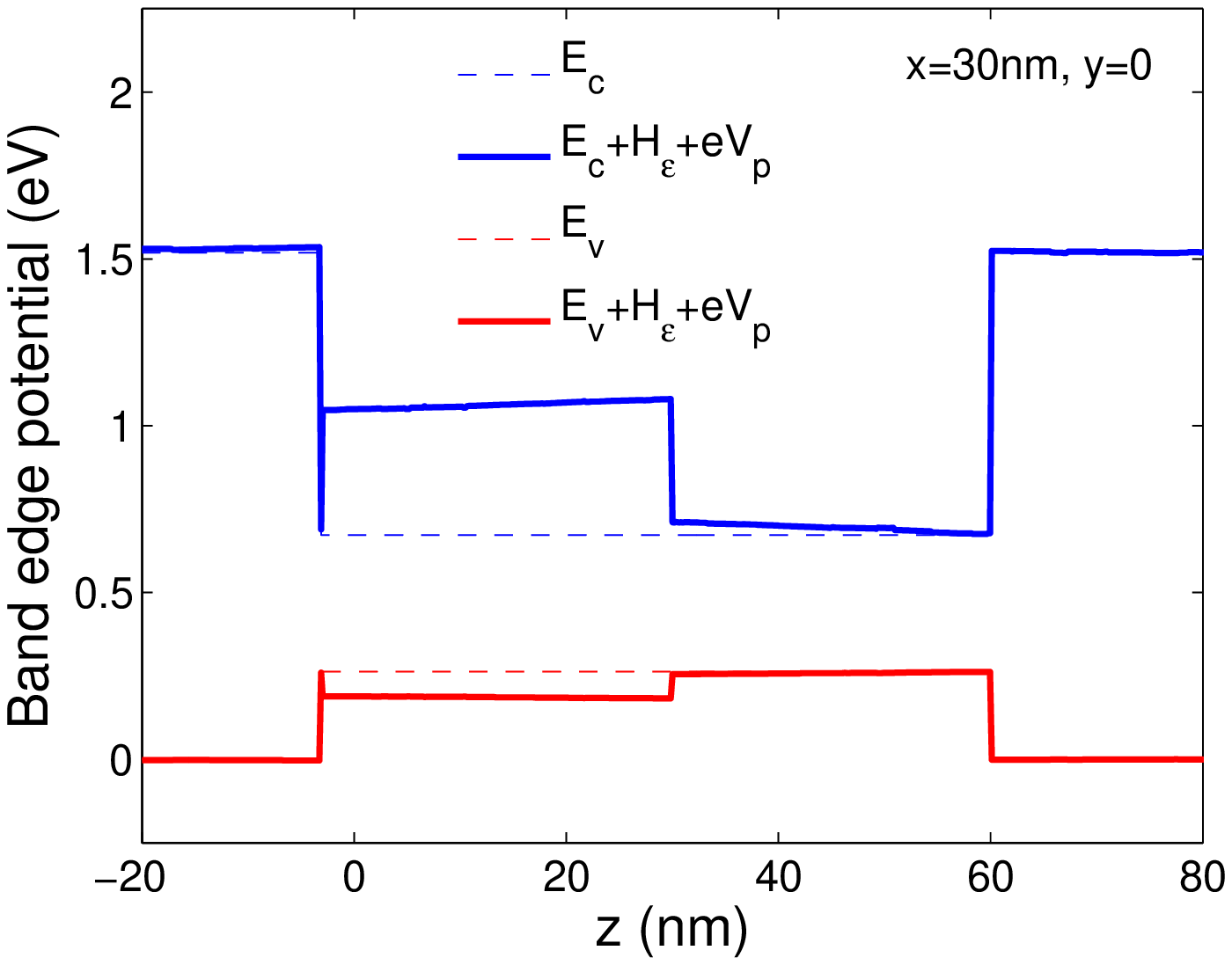, height=4.5cm,width=12.2cm}}
%\end{tabular}
%\vspace{-0.8cm}
\caption{The band edge potential with the full nonlinear strain model (x=30, y=0).}
\end{center}
%\vspace{-0.1cm}
\end{figure}
%\vspace{-1cm}

\begin{figure}
\begin{center}
{\epsfig{file=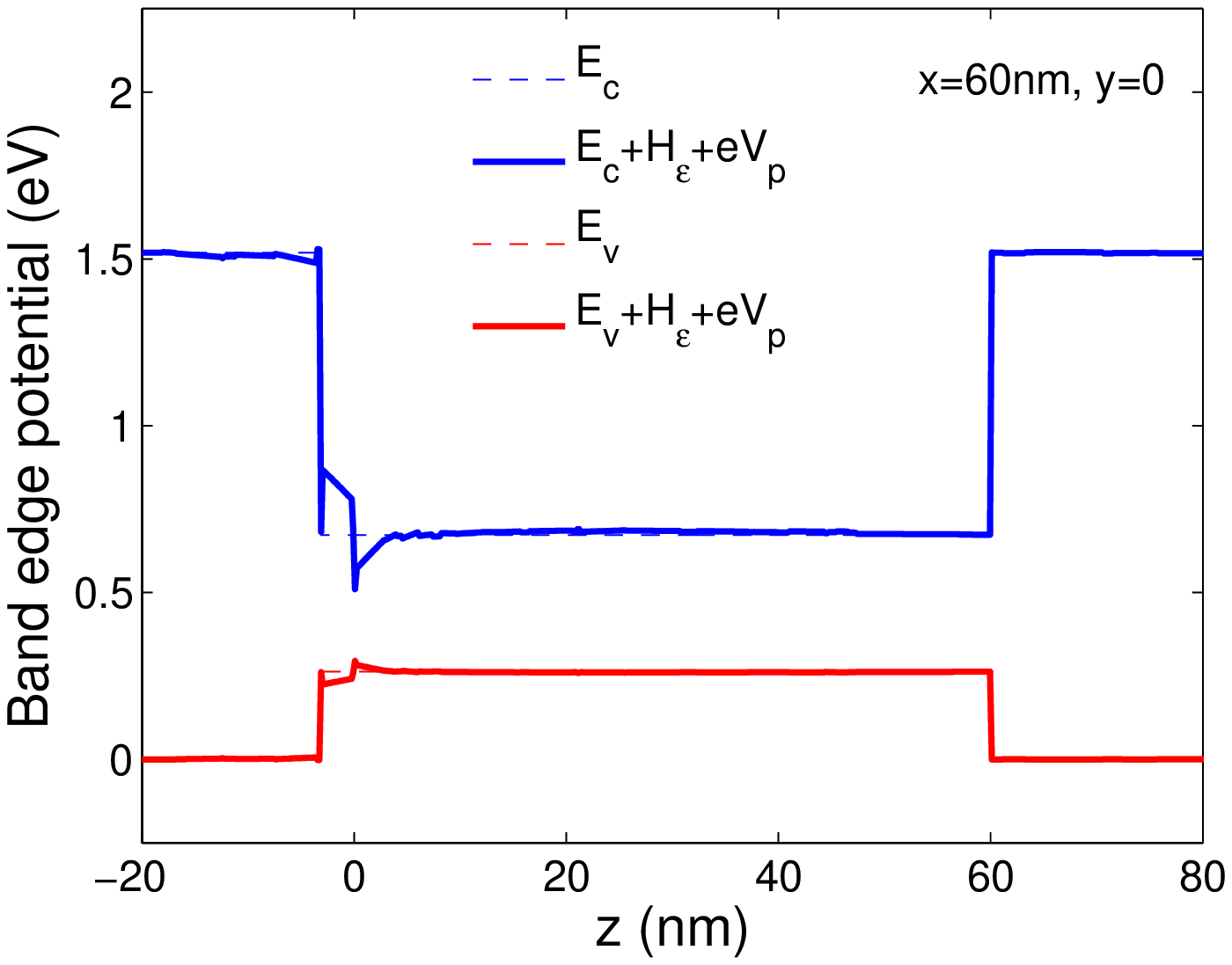, height=4.5cm, width=12.2cm}}
%\end{tabular}
%\vspace{-0.8cm}
\caption{The band edge potential with the full nonlinear strain model (x=60, y=0).}
\end{center}
%\vspace{-0.1cm}
\end{figure}
%\vspace{-1cm}

\begin{table}[h]
\begin{tabular}{|c|c|c|c|c|c|}
\hline
{\rm Eigenstate $\#$} & {\rm Lin/strain} & {\rm Lin/strain+piezo} & {\rm Case I} &
{\rm
Case II} & {\rm Case III} \\
 \hline
 1 & 0.7122 & 0.6968 & 0.6811 & 0.6913 & 0.6767 \\
 \hline
 2 & 0.8345 & 0.8084 & 0.8008 & 0.8070 & 0.7994 \\
 \hline
 3 & 0.8404 & 0.8115 & 0.8046 & 0.8095 & 0.8027 \\
 \hline
 4 & 0.8511 & 0.8272 & 0.8248 & 0.8267 & 0.8244 \\
 \hline
 5 & 0.8658 & 0.8350 & 0.8321 & 0.8345 & 0.8316 \\
\hline
\end{tabular}
\caption{The influence of strain, piezoeffects, and nonlinear contributions on eigenstates of
the structure.}
\end{table}

Finally, we calculate the bandstructure with our finite element method code and compare the
band edge potential due to the above three nonlinear strain models in the presence of
piezoelectricity. The difference for each of these three cases of band edge potentials with
respect to that of the linear case is computed and plotted along the z-axis for (a)
$x=0,y=0$, (b) $x=30 {\rm nm}, y=0$, and (c) $x=60 {\rm nm}, y=0$ in Figs. 13--15,
respectively. It is clear from the plots that the fully nonlinear strain model (Case III)
captures the large inhomogeneous strain-induced band edge potential as a cumulative effect of
the potentials for Cases I and II.  The band edge potentials along the presented [001]
direction (z-axis) for each of these three cases have been computed with the fully coupled
model.

\begin{figure}
\begin{center}
{\epsfig{file=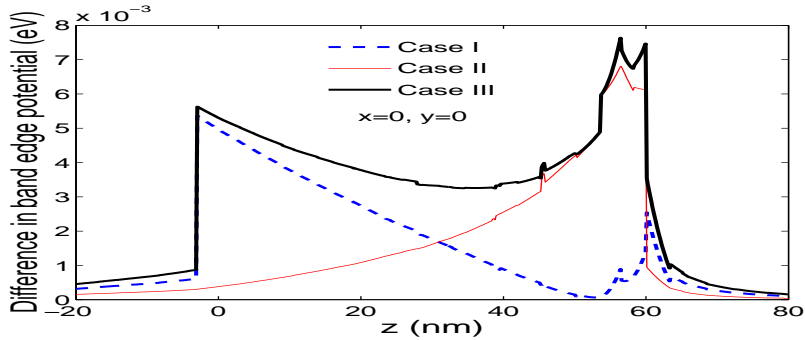, height=4.5cm, width=12.2cm}}
%\end{tabular}
%\vspace{-0.8cm}
\caption{Quantifying the difference between linear and nonlinear models in the presence of
piezoeffect (x=0, y=0).}
\end{center}
%\vspace{-0.1cm}
\end{figure}
%\vspace{-1cm}

\begin{figure}
\begin{center}
{\epsfig{file=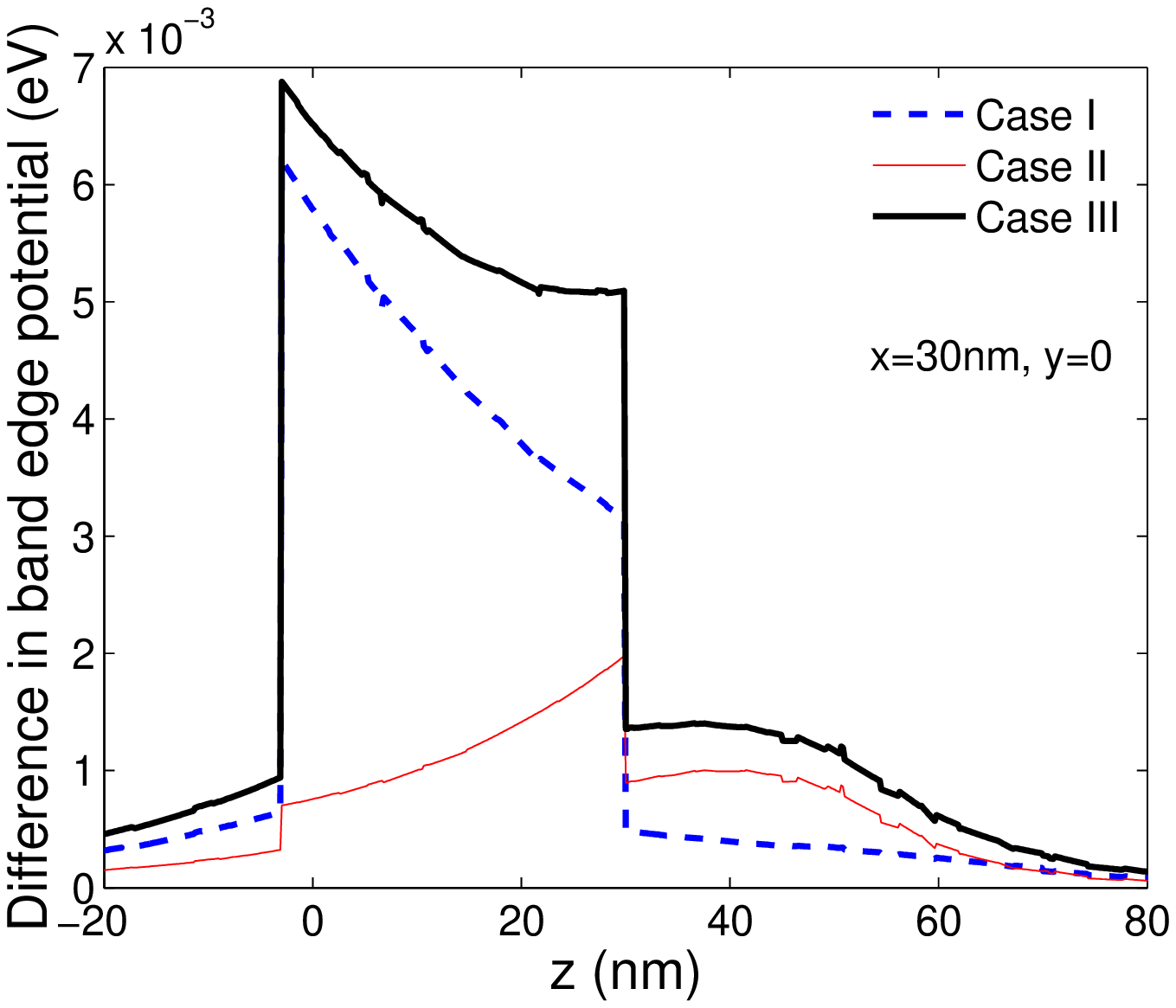, height=4.5cm, width=12.2cm}}
%\end{tabular}
%\vspace{-0.8cm}
\caption{Quantifying the difference between linear and nonlinear models in the presence of
piezoeffect (x=30, y=0).}
\end{center}
%\vspace{-0.1cm}
\end{figure}
%\vspace{-1cm}

\begin{figure}
\begin{center}
{\epsfig{file=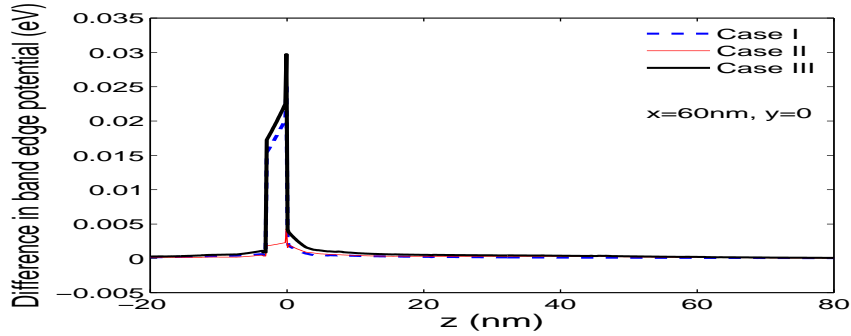, height=4.5cm, width=12.2cm}}
%\end{tabular}
%\vspace{-0.8cm}
\caption{Quantifying the difference between linear and nonlinear models in the presence of
piezoeffect (x=60, y=0).}
\end{center}
%\vspace{-0.1cm}
\end{figure}
%\vspace{-1cm}

%%%%%%%%%%%%%%%%%%%%%%
\section{Conclusions}

In this contribution, we analyzed two fundamental approaches in
bridging the scales in mathematical models for the description of
optoelectromechanical properties of nanostructures. We demonstrated
that the conventional application of linear models to the analysis
of properties of nanostructures in bandstructure engineering could
be inadequate. By accounting consistently for the piezoelectric
effect, we considered three nonlinear strain models and analyzed
contributions of quadratic nonlinear terms induced by the
deformation gradient in the growth direction, as well as in the
plane normal to that direction. The core of our model was based on
the Shrodinger-Poisson system which was presented in the variational
form and implemented with finite element methodology. In this
framework, we presented also the full nonlinear strain model and
quantify the differences between the conventional linear and
developed here
 nonlinear models for bandstructure calculations. Generalizations of
the existing models and the examples provided for quantum dot
nanostructures emphasized the importance of coupled effects in
predicting optoelectromechanical properties of such structures.

% \appendix

% \section{}
% \label{}

\section*{Acknowledgments}

This work was supported by NSERC. The authors thank Profs M.
Willatzen (Denmark), L.C. Lew Yan Voon (USA), and Dr B. Lassen
(Sweden) for fruitful discussions on LDSN modelling issues and the
anonymous referees for a number of useful suggestions incorporated
in the final version of this paper.


\begin{thebibliography}{00}

\bibitem{Grundmann1995} Grundmann M, Stier O, Bimberg D. InAs/GaAs pyramidal quantum dots:
strain distribution, optical phonons, and electronic structure. Phys. Rev. B 1995; 52(16):
11969.

\bibitem{Andreev2000} Andreev AD, O'Reilly EP. Theory of the electronic structure of GaN/AlN
hexagonal quantum dots. Phys. Rev. B 2000; 62(23): 15851--15870.

\bibitem{Johnson2003} Johnson HT, Bose R. Nanoindentation effect on the optical properties of
self-assembled quantum dots. J. Mech. Phys. Solids. 2003; 51: 2085--2104.

\bibitem{Fonoberov2003} Fonoberov VA, Balandin AA. Excitonic properties of strained wurtzite
and zinc-blende $GaN/Al_xGa_{1-x}N$ quantum dots. J. Appl. Phys. 2003; 94(11): 7178-7186.

\bibitem{Kuo2006} Kuo MK, Lin TR, Hong KB et al. Two-step strain analysis of self-assembled
InAs/GaAs quantum dots. Semicond. Sci. Technol. 2006; 21: 626--632.

\bibitem{Pan2002} Pan E. Elastic and piezoelectric fields around a quantum dot: fully coupled
or semicoupled model? J. Appl. Phys. 2002; 91(6): 3785--3796.

\bibitem{Jogai2003} Jogai B, Albrecht, JD, Pan E. Effect of electromechanical coupling on the
strainin AlGaN/GaN HFETs. J. Appl. Phys. 2003; 94: 3984--3989.

\bibitem{Melnik2004} Melnik RVN, Willatzen M. Bandstructures of conical quantum dots with wetting layers.
Nanotechnology. 2004; 15: 1--8.


\bibitem{Willatzen2004MCS} Willatzen M., Melnik RVN, Galeriu C., and Lew Yan Voon LC.
Quantum confinement phenomena in nanowire superlattice structures.
Mathematics and Computers in Simulation. 2004; 65(4-5): 385--397.



\bibitem{Bass1997} Bass F.G. and Bulgakov A.A. Kinetic and Electrodynamic Phenomena in
Classical and Quantum Semiconductor Superlattices. Nova Science
Publisher. 1997.



\bibitem{Melnik2000MS} Melnik RVN, He H. Relaxation-time approximations of quasi-hydrodynamic type in
semiconductor device modelling.  Modelling and Simulation in
Materials Science and Engineering. 2000; 8(2): 133-149.

\bibitem{Melnik2000MCS} Melnik RVN, He H. Quasi-hydrodynamic modelling and computer simulation of coupled
thermo-electrical processes in semiconductors. Mathematics and
Computers in Simulation.2000; 52(3-4): 273-287.


\bibitem{Rudan2005} Rudan M, Gnani E, Reggiani
S, Baccarani G. A coherent extension of the transport equations in
semiconductors incorporating the quantum correction. IEEE
Transcations on Nanotechnology. 2005; 4(5): 495-509.

\bibitem{Ipatova1993} Ipatova IP, Malyshkin VG, Shchukin VA. On
spinoidal decomposition in ellastically anisotropic epitaxial films
of III-V semiconductor alloys. J. Appl. Phys. 1993; 74(12):
7198--7210.


\bibitem{Bir1974} Bir GL, Pikus GE. Symmetry and strain-induced effects in semiconductors.
New York: Wiley, 1974.

\bibitem{Melnik2004ZPiezo} Melnik RVN, Zotsenko KN. Mixed electrostatic waves and CFL stability conditions
in computational piezoelectricity. Appl. Numer. 2004; 48(1): 41--62.


\bibitem{Melnik2004Z} Melnik RVN, Zotsenko KN. Finite element analysis of coupled electronic
states in quantum dot nanostructures. Model. Simul. Mater. Sci. Eng. 2004; 12(3): 465--477.


\bibitem{Melnik1998} Melnik RVN, Melnik KN. A note on the class of weakly coupled problems of
non-stationary piezoelectricity. Commun. Numer. Meth. Engnr. 1998; 14: 839--847.

\bibitem{Melnik2000} Melnik RVN. Generalized solutions, discrete models and energy estimates
for a 2D problem of coupled field theory. Appl. Math. Comput. 2000; 107: 27--55.


\bibitem{Voss2006} Voss H. Numerical calculation of the electronic structure for
three-dimensional quantum dots. Comp. Phys. Comm. 2006; 174: 441--446.


\bibitem{Melnik2000H} Melnik RVN, He H. Modelling nonlocal processes in
semiconductor devices with exponential difference schemes. J. Eng. Math. 2000; 38: 233--263.

\bibitem{Bourgade2006} Bourgade JP, Degond P, Mehats F, Ringhofer C. On quantum extensions
to classical spherical harmonics expansion/Fokker-Planck models.  J. Math. Phys. 2006; 47
(4): 043302.

\bibitem{Lassen2004} Lassen B, Voon LCLY, Willatzen M, Melnik RVN. Exact envelope-function
theory versus symmetrized Hamiltonian for quantum wires: a
comparison. Solid State Communications. 2004; 132(3-4): 141--149.


\bibitem{Bastard1988} Bastard G. Wave mechanics applied to semiconductor
 heterostructures. New York: Halsted Press (a division of John Wiley
 \& Sons), 1988.

\bibitem{Davies1998} Davies JH. The physics of low-dimensional
semiconductors. Cambridge: Cambridge University Press, 1998.

\bibitem{Singh1993} Singh J. Physics of semiconductors and their
heterostructures. New York: McGraw-Hill, 1993.

%%%%%%%%%%%%%%%%%%%%%


% notes:

% \bibitem{label2}
% If there is a note, it should come last:
% \bibitem{label3} \note
% \end{subbibitems}

\end{thebibliography}
\end{document}